\documentclass[10pt,final,twocolumn]{IEEEtran}
\renewcommand{\baselinestretch}{0.99}
\usepackage{hyperref}
\usepackage{cite}
\usepackage{amsmath,amsthm,amssymb,amsfonts,steinmetz}
\usepackage{graphicx,flushend}
\usepackage[usenames,dvipsnames]{color}

\def\BibTeX{{\rm B\kern-.05em{\sc i\kern-.025em b}\kern-.08em
    T\kern-.1667em\lower.7ex\hbox{E}\kern-.125emX}}

\usepackage{tikz}
\usetikzlibrary{calc}
\makeatletter
\newcommand{\gettikzxy}[3]{%
  \tikz@scan@one@point\pgfutil@firstofone#1\relax
  \edef#2{\the\pgf@x}%
  \edef#3{\the\pgf@y}%
}
\makeatother
\usepackage{color,soul}
\usepackage{comment}
\usepackage{xcolor}
\usepackage{algorithm,algorithmic}
\usepackage{subcaption}

\renewcommand{\a}{\mathbf{a}}

\newcommand{\f}{\mathbf{f}}
\newcommand{\g}{\mathbf{g}}
\newcommand{\h}{\mathbf{h}}

\newcommand{\n}{\mathbf{n}}

\newcommand{\p}{\mathbf{p}}
\newcommand{\q}{\mathbf{q}}

\newcommand{\s}{\mathbf{s}}

\renewcommand{\u}{\mathbf{u}}
\renewcommand{\v}{\mathbf{v}}
\newcommand{\w}{\mathbf{w}}
\newcommand{\x}{\mathbf{x}}
\newcommand{\y}{\mathbf{y}}

\newcommand{\A}{\mathbf{A}}

\newcommand{\D}{\mathbf{D}}

\newcommand{\F}{\mathbf{F}}
\newcommand{\G}{\mathbf{G}}
\renewcommand{\H}{\mathbf{H}}
\newcommand{\I}{\mathbf{I}}
\newcommand{\J}{\mathbf{J}}

\renewcommand{\L}{\mathbf{L}}

\newcommand{\N}{\mathbf{N}}

\renewcommand{\P}{\mathbf{P}}
\newcommand{\Q}{\mathbf{Q}}

\newcommand{\U}{\mathbf{U}}

\newcommand{\W}{\mathbf{W}}

\newcommand{\Y}{\mathbf{Y}}

\newcommand{\Compl}{\mbox{$\mathbb{C}$}}

\renewcommand{\Im}{\mathrm{Im}}

\renewcommand{\Re}{\mathrm{Re}}

\DeclareMathAlphabet\mathbfcal{OMS}{cmsy}{b}{n}

\begin{document}
\setlength{\parindent}{10pt}
\title{Extremely Large Full Duplex MIMO\\ for Simultaneous Downlink Communications\\ and Monostatic Sensing at Sub-THz Frequencies}
\author{George C. Alexandropoulos,~\IEEEmembership{Senior~Member,~IEEE}, and Ioannis Gavras,~\IEEEmembership{Student~Member,~IEEE}
\thanks{Part of this work has been presented in IEEE SPAWC, Lucca, Italy, September 2024~\cite{spawc2024}.}
\thanks{This work has been supported by the SNS JU projects TERRAMETA and 6G-DISAC under the EU's Horizon Europe research and innovation programme under Grant Agreement numbers 101097101 and 101139130, respectively. TERRAMETA also includes top-up funding by UKRI under the UK government's Horizon Europe funding guarantee.}
\thanks{The authors are with the Department of Informatics and Telecommunications, National and Kapodistrian University of Athens, 16122 Athens, Greece. G. C. Alexandropoulos is also with the Department of Electrical and Computer Engineering, University of Illinois Chicago, IL 60601, USA (e-mails: \{giannisgav, alexandg\}@di.uoa.gr).}
}

\maketitle
\begin{abstract}
The in-band Full Duplex (FD) technology is lately gaining attention as an enabler for the emerging paradigm of Integrated Sensing and Communications (ISAC), which envisions seamless integration of sensing mechanisms for unconnected entities into next generation wireless networks. In this paper, we present an FD Multiple-Input Multiple-Output (MIMO) system with extremely large antenna arrays at its transceiver module, which is optimized, considering two emerging analog beamforming architectures, for simultaneous DownLink (DL) communications and monostatic-type sensing intended at the sub-THz frequencies, with the latter operation relying on received reflections of the transmitted information-bearing signals. A novel optimization framework for the joint design of the analog and digital transmit beamforming, analog receive combining, and the digital canceler for the self-interference signal is devised with the objective to maximize the achievable DL rate, while meeting a predefined threshold for the position error bound for the unknown three-dimensional parameters of a passive target. Capitalizing on the distinctive features of the beamforming architectures with fully-connected networks of phase shifters and partially-connected arrays of metamaterials, two ISAC designs are presented. Our simulation results showcase the superiority of both proposed designs over state-of-the-art schemes, highlighting the role of various system parameters in the trade-off between the communication and sensing functionalities.
\end{abstract}

\begin{IEEEkeywords}
Integrated sensing and communications, full duplex, extremely large MIMO, metasurfaces, THz, optimization.
\end{IEEEkeywords}

\section{Introduction}
Integrated Sensing And Communications (ISAC) is an already widely accepted emerging feature for the upcoming sixth Generation (6G) of wireless networks\cite{mishra2019toward,liu2022integrated}, which is envisioned as a new standalone usage scenario by ITU's radiocommunication sector for the international mobile telecommunications systems of $2030$~\cite{ITU-R}. Channel modeling and use cases for ISAC are currently being studied by 3GPP's technical specification group on the radio access network for the upcoming Release $19$~\cite{3GPP_R19}, while ETSI has recently launched an industry specification group dedicated to also specify key performance indicators for ISAC and their evaluation methodology, study relevant radio access technologies for sensing together with the necessary system architecture modifications for the different sensing types (e.g., monostatic, multistatic, or combinations thereof), as well as define mechanisms to meet security and privacy requirements for ISAC~\cite{ETSI_ISAC}. In parallel, both industry and academia are investigating enabling ISAC technologies~\cite{6G-DISAC_mag,gonzalez2024integrated,wang2024integration}, including eXtremely Large (XL) Multiple-Input Multiple-Output (MIMO) systems~\cite{XLMIMO_tutorial} operating up to at THz frequencies~\cite{THs_loc_survey} to profit from enhanced angular and range resolution, Reconfigurable Intelligent Surfaces (RISs) that enable programmable correlations between sensing and communication subspaces for ISAC boosting~\cite{RIS_ISAC}, as well as Full Duplex (FD) radios~\cite{FD_JSAC_Survey} that have been lately considered for simultaneous communications and monostatic-type sensing, with both functionalities relying solely on the transmission and reflection of information-bearing signals~\cite{FD_ISAC_SPM}.

Current ISAC studies offer a large collection of metrics for the combination of communications and sensing, resulting in an extensive pallet of relevant trade-offs. For example, the authors in~\cite{fei2024revealing} used a unified Kullback-Leibler divergence metric to explore the Pareto bound of a monostatic ISAC system based on a multi-antenna base station. In~\cite{liu2021cramer} and~\cite{wang2022partially}, using the Cram\'{e}r-Rao bound (CRB) as the sensing performance metric, ISAC BeamForming (BF) designs for both the Fully Digital Antenna (FDA) array architecture as well as for one based on hybrid Analog and Digital (A/D) beamformers were presented. It was showcased that this metric yields superior estimation performance compared to ISAC designs focusing on the optimization of the beampatterns at the transmitter (TX). The authors in~\cite{wang2024joint} proposed a joint transceiver optimization design for millimeter wave and THz multi-user ISAC scenarios, considering the trade-off between communication's sum-rate metric and the signal-to-clutter-and-noise ratio. The fundamental trade-off between communication's rate and the CRB for sensing in Gaussian channels was analyzed in~\cite{xiong2023fundamental}. The authors in~\cite{furkan2024fundamental} investigated the time-frequency and spatial trade-offs, originating respectively from the choice of modulation order for random data and the design of BF strategies, in monostatic ISAC systems with Orthogonal Frequency-Division Multiplexing (OFDM). In~\cite{tian2024performance}, the trade-off between the localization and communication functionalities for a MIMO setup incorporating backscatter communications and performing bistatic-type sensing was analytically studied.

The simultaneous transmission and reception capabilities of FD systems are lately inspiring wireless communications researchers as an efficient means to realize simultaneous bidirectional links, one for data communications and the other for sensing~\cite{barneto2019full,liyanaarachchi2021optimized}. Among the first works that presented ISAC frameworks for sensing-assisted MIMO communications belong~\cite{MultiuserComms-CE2020, SingleuserComms-CE2020,Direction-Aided2020} that, capitalizing on the multi-antenna transceiver architectures in~\cite{FD_MIMO_VTM2022}, deployed the reception capability of a single FD MIMO node in time-division duplexing setups to enable estimation of communication channel parameters simultaneously with the transmission of data signals. An FD-enabled multi-beam ISAC system, where the TX and Receive (RX) beamformers were optimized to realize multiple beams for both communications and sensing, was presented in~\cite{barneto2020beamforming}. The received reflections at an FD node's RX from the data signals transmitted by its TX in the DownLink (DL) were used in~\cite{liyanaarachchi2021joint} to estimate the range and angle profiles corresponding to multiple targets lying in the node's vicinity. In~\cite{Atiq_ISAC_2022,Islam_2022_ISAC,Asilomar_FD_HMIMO_2023}, FD MIMO ISAC schemes that perform Downlink (DL) data communications towards to a single and multiple User Equipments (UEs), while simultaneously estimating parameters related to radar targets, were designed. Precoding designs based on the eigendecomposition of a composite channel matrix including the matrices for DL communications and Self-Interference (SI), while taking into account a radar-oriented gain, were presented in~\cite{bayraktar2023self,bayraktar2023hybrid} for a millimeter wave FD MIMO system with hybrid A/D TX and RX beamformers. An optimized synergetic ISAC operation of an FD MIMO node with an RIS placed in the former's near-field region was proposed in~\cite{FD_RIS_EUSIPCO2023}. Targeting simultaneous near-field DL communications and monostatic-type sensing at THz frequencies, \cite{FD_HMIMO_2023} introduced an XL in-band FD MIMO transceiver architecture with TX and RX Dynamic Metasurface Antennas (DMAs)~\cite{Shlezinger2021Dynamic}, and presented a joint design of its A/D BF and digital SI cancellation parameters with the objective to maximize the Signal-to-Noise Ratios (SNRs) for both functionalities. The sensing operation of this ISAC approach was recently extended in~\cite{gavras2024joint} to enable tracking of the target's spatial parameters.  

In this paper, we present a novel XL in-band FD MIMO system for integrated DL communications and monostatic-type sensing, with the latter operation leveraging the reflections of the DL data signals from radar targets lying in the FD node's vicinity. We consider a near-field channel model for sub-THz frequencies and focus on an efficient narrowband framework where both functionalities are realized simultaneously. Differently from the FD-enabled ISAC state-of-the-art schemes in~\cite{barneto2020beamforming,liyanaarachchi2021joint,Atiq_ISAC_2022,Islam_2022_ISAC,Asilomar_FD_HMIMO_2023,bayraktar2023self,bayraktar2023hybrid,FD_RIS_EUSIPCO2023,FD_HMIMO_2023,gavras2024joint}, we devise a novel design for the TX and RX A/D beamformers and the digital SI cancellation at the FD node, with the objective to maximize the performance of DL communications while guaranteeing an estimation-oriented performance bound threshold. The main technical contributions of the paper are summarized as follows.
\begin{itemize}
    \item Leveraging \cite{FD_ISAC_SPM}'s recent interpretation of the SI channel for FD-enable MIMO ISAC systems, we present a convenient model for the received signal at the RX side of the XL FD MIMO node elaborating on the acquisition of its different components.
    \item A novel analysis for the Position Error Bound (PEB) with respect to a target's three-Dimensional (3D) spatial parameters, considering two analog BF architectures for the TX and RX sides of the XL FD MIMO node, one incorporating fully-connected networks of Phase Shifters (PSs) and the other based on partially-connected arrays of metamaterials (including conventional partially-connected antenna arrays as a special case), is presented.
    \item We formulate a novel optimization framework for the joint design of the A/D TX BF matrix/vector, the analog RX BF matrix, and the digital SI cancellation matrix at the XL FD MIMO node, targeting maximization of the achievable rate towards a UE in the DL direction, while meeting a PEB threshold for the unknown target's parameters. Leveraging the distinctive features of the considered TX/RX analog BF architectures, two efficient solutions for the overall system design are presented.  
    \item Our extensive performance evaluation results for various settings of the system parameters showcase the superiority of both proposed ISAC designs over state-of-the-art schemes, and highlight the potential of the DMA array architecture for the proposed XL FD MIMO system. The role of the scaling of our system's core components and that of the residual SI level in the trade-off between the communication and sensing functionalities are also numerically assessed via evaluating the Pareto boundary. 
\end{itemize}

A preliminary version of our ISAC framework has been presented very recently in~\cite{spawc2024}, however, in this paper, we derive the exact PEB of the unknown target, considering two emerging analog BF architectures, and then incorporate it in our design optimization formulation, which is solved via new approaches, one per BF architecture. Herein, we also investigate the Pareto boundary between the communication and sensing functionalities and include additional simulation results for diverse settings of critical system parameters. 

The remainder of this paper is organized as follows. Section~\ref{Sec: System} includes the models for the XL FD MIMO system, the sub-THz wireless channel, and the received signal for both analog BF architectures. Section~\ref{Sec: Opt} presents our FD-enabled ISAC optimization formulation together with its solution for both analog BF architectures, while Section~\ref{Sec: Numerical} provides an extensive description of our simulation results and comparisons with state-of-the-art FD-enabled ISAC schemes. Finally, Section~\ref{ref:Conclusion} contains the concluding remarks of the paper.

\textit{Notations:}
Vectors and matrices are represented by boldface lowercase and uppercase letters, respectively. The transpose, Hermitian transpose, inverse, and Moore-Penrose pseudoinverse of $\mathbf{A}$ are denoted as $\mathbf{A}^{\rm T}$, $\mathbf{A}^{\rm H}$, $\mathbf{A}^{-1}$, and $\mathbf{A}^{\rm \dagger}$ respectively. $\mathbf{I}_{n}$, $\mathbf{0}_{n}$, and $\boldsymbol{1}_n$ ($n\geq2$) indicate the $n\times n$ identity, zeros' matrices, and ones' column vector, respectively. $[\mathbf{A}]_{i,j}$ is the $(i,j)$-th element of $\mathbf{A}$, whereas $i:j$ as a matrix row/column index indicates the respective $i$-th till the $j$-th elements. $\|\mathbf{A}\|$ and $\|\mathbf{A}\|_{\rm F}$ represent $\mathbf{A}$'s Euclidean and Frobenious norms, respectively. $|a|$, ${\rm arg}(a)$, and $\Re\{a\}$ are respectively the amplitude, phase angle, and real part of complex scalar $a$, $\mathbb{C}$ is the complex number set, and $\jmath\triangleq\sqrt{-1}$ is the imaginary unit. $\mathbb{E}\{\cdot\}$ is the expectation operator and $\mathbf{x}\sim\mathcal{CN}(\mathbf{a},\mathbf{A})$ represents a complex Gaussian random vector with mean $\mathbf{a}$ and covariance matrix $\mathbf{A}$. ${\rm vec(\cdot)}$ is the vectorization operation with ${\rm unvec(\cdot)}$ denoting the inverse, and $\angle\cdot$ provides the phases of the elements of a complex matrix. Finally, $\circ$ is the element-wise multiplication operation and $\otimes$ gives the Kroneker product. 

\section{System and Channel Models}\label{Sec: System}
\begin{figure}
\centering
        \includegraphics[scale=1]{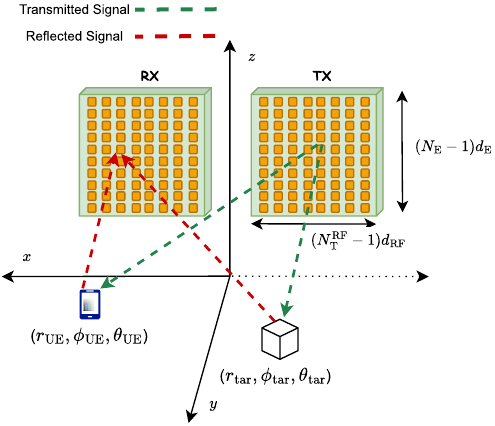}
        \caption{\small{The proposed XL in-band FD MIMO system for simultaneous DL data communications and monostatic-type sensing.}}
        \label{fig: system}
\end{figure}

Consider an XL in-band FD MIMO system comprising one TX and one RX Uniform Planar Arrays (UPAs), both being capable of realizing hybrid A/D signal processing~\cite{Vishwanath_2020,alexandropoulos2020full}. This system aims to establish DL communications with a single-antenna UE, while concurrently performing monostatic sensing of a passive radar target, solely through the reception of the reflections of the DL data signal from that target. As depicted in Fig.~\ref{fig: system}, the system's TX and RX antenna panels, composed respectively of $N_{\rm T} \triangleq N^{\rm RF}_{\rm T}N_{\rm E}$ and $N_{\rm R} \triangleq N^{\rm RF}_{\rm R}N_{\rm E}$ elements, are placed in the $xz$-plane. It is assumed that the $N_{\rm E}$ uniformly spaced antenna elements at each TX (RX) UPA column are attached to the same TX (RX) Radio Frequency (RF) chain~\cite{alexandropoulos2017joint} (i.e., there exist in total $N^{\rm RF}_{\rm T}$ and $N^{\rm RF}_{\rm R}$ TX and RX RF chains, respectively), and that the $N^{\rm RF}_{\rm T}$ ($N^{\rm RF}_{\rm R}$) antennas at each TX (RX) UPA's row are uniformly spaced.  

The TX portion of the considered FD node possesses the UE's complex-valued symbol $s$, which is first precoded via the digital BF vector $\v\in\Compl^{N^{\rm RF}_{\rm T}\times 1}$. Before transmission in the DL, the digitally precoded symbols are analog processed via the analog TX BF matrix $\W_{\rm TX}\in\Compl^{N_{\rm T}\times N^{\rm RF}_{\rm T}}$, resulting in the $N_{\rm T}$-element transmitted signal $\x\triangleq\W_{\rm TX}\v s$, which is assumed power limited such that $\mathbb{E}\{\|\W_{\rm TX}\v s\|^2\}\leq P_{\rm{\max}}$, where $P_{\rm{\max}}$ represents the maximum transmission power. The RX part of the FD node is tasked to efficiently sense the target lying in its vicinity, as illustrated in Fig.~\ref{fig: system}. Upon reception, the reflections of the DL data signals from the target are first processed in the analog domain by the RX BF matrix $\W_{\rm RX}\in\Compl^{N_{\rm R}\times N^{\rm RF}_{\rm R}}$, and then, in the digital domain, by the matrix $\D\in\Compl^{N_{\rm R}^{\rm RF}\times N_{\rm T}^{\rm RF}}$ to enable efficient estimation of the target's spatial parameters.  

\subsection{Analog BF Architectures}\label{subsec:Trans}
In this paper, we consider the following two TX/RX analog BF architectures for the XL in-band FD MIMO system, whose tunable parameters will be later on optimized for the proposed simultaneous DL data communications and monostatic-type sensing functionality.

\subsubsection{Fully-Connected Network of Phase Shifters (FCPS)}\label{Sec:PS_ARC}
According to this architecture, the output (input) of each TX (RX) RF chain is connected with each of the antenna elements via a dedicated PS of tunable weight~\cite{MRV17_MIMO_Survey}. Let $w^{\rm TX}_{i,n}$ ($w^{\rm RX}_{i,n}$) denote the analog weight of the PS associated with each $n$-th antenna element of each $i$-th TX (RX) RF chain, where $n=1,2,\ldots,N_{\rm T} (N_{\rm R})$ and $i=1,2,\ldots,N_{\rm T}^{\rm RF} (N_{\rm R}^{\rm RF})$. These weights are assumed to belong in the codebook $\mathcal{W}_{\rm PS}$, i.e.:
\begin{align}\label{eq: Hybrid_code}
    w^{\rm TX}_{i,n}, w^{\rm RX}_{i,n} \in \mathcal{W}_{\rm PS}\triangleq \left\{e^{\jmath\phi}\Big|\phi\in\left[-\frac{\pi}{2},\frac{\pi}{2}\right]\right\}.
\end{align}
Note that the analog weight $w^{\rm TX}_{i,n}$ represents the $(i,n)$-th element of the analog TX BF matrix $\W_{\rm TX}$; similarly does the weight $w^{\rm RX}_{i,n}$ for the analog RX BF matrix $\W_{\rm RX}$.

\subsubsection{Partially-Connected Arrays of Metamaterials (PCM)}\label{Sec:DMA_ARC}
DMAs have been recently considered for efficiently realizing XL MIMO transceivers~\cite{HMIMO_survey}, facilitating the packaging of extremely large numbers of sub-wavelength-spaced metamaterials in compact apertures~\cite{Shlezinger2021Dynamic}. These metasurface-based front-ends consist of multiple single-RF-fed microstrips of linearly equally spaced metamaterials of dynamically tunable responses, which are capable of effectively implementing analog TX/RX BF. According to this architecture, each TX (RX) microstrip is driven by a single TX (RX) RF chain. The input-output relationship in a TX (RX) DMA with $N_{\rm T}$ ($N_{\rm R}$) metamaterials can be represented by the $N_{\rm T}\times N_{\rm T}$ ($N_{\rm R}\times N_{\rm R}$) diagonal matrix $\P_{\rm TX}$ ($\P_{\rm RX}$), whose elements model signal propagation inside all TX (RX) microstrips. The diagonal elements of $\P_{\rm TX}$ ($\P_{\rm RX}$) are defined $\forall i=1,2,\ldots,N^{\rm RF}_{\rm T} (N^{\rm RF}_{\rm R})$ and $\forall n = 1,2,\ldots,N_{\rm E}$, using $i_n\triangleq (i-1)N_{\rm E}+n$, as~\cite{Xu_DMA_2022}:
\begin{align}\label{eq: TX_Sig_Prop}
    [\P_{\rm TX}]_{i_n,i_n}, [\P_{\rm RX}]_{i_n,i_n} \triangleq \exp{(-\rho_{i,n}(\alpha_i + \jmath\beta_i))},
\end{align} 
where $\alpha_i$ is the waveguide attenuation coefficient, $\beta_i$ is the wavenumber, and $\rho_{i,n}$ denotes the location of the $n$-th element in the $i$-th TX (RX) microstrip. Let $\widetilde{w}^{\rm TX}_{i,n}$ ($\widetilde{w}^{\rm RX}_{i,n}$) denote the adjustable response weight of each $n$-th metamaterial of each $i$-th TX (RX) microstrip. These weights are usually assumed to follow a Lorentzian-constrained model belonging to the codebook $\mathcal{W}_{\rm L}$, yielding~\cite{Xu_DMA_2022}:
\begin{align}\label{eq: DMA_code}
    \widetilde{w}^{\rm TX}_{i,n}, \widetilde{w}^{\rm RX}_{i,n} \in \mathcal{W}_{\rm L}\triangleq \left\{\frac{\jmath+e^{\jmath\phi}}{2}\Big|\phi\in\left[-\frac{\pi}{2},\frac{\pi}{2}\right]\right\}.
\end{align}
Using this definition, the TX DMA's analog weight matrix $\widetilde{\W}_{\rm TX}\in\mathbb{C}^{N_{\rm T}\times N^{\rm RF}_{\rm T}}$ is defined for $i=j$ as $[\widetilde{\W}_{\rm TX}]_{i_n,j}\triangleq\widetilde{w}^{\rm TX}_{i,n}$ and for $i\neq j$ as $[\widetilde{\W}_{\rm TX}]_{i_n,j}\triangleq0$. Consequently, the $N_{\rm T}\times N_{\rm T}^{\rm RF}$ analog TX BF matrix is obtained as $\W_{\rm TX}=\P_{\rm TX}\widetilde{\W}_{\rm TX}$, where signal propagation within each microstrip has been taken into account. Similarly, we obtain the analog weight matrix $\widetilde{\W}_{\rm RX}\in\mathbb{C}^{N_{\rm R}\times N_{\rm R}^{\rm RF}}$ of the RX DMA, and then, the $N_{\rm R}\times N_{\rm R}^{\rm RF}$ analog RX BF matrix $\W_{\rm RX}=\P_{\rm RX}\widetilde{\W}_{\rm RX}$. 

\subsection{Channel Model}
The $1\times N_{\rm T}$ complex-valued DL sub-THz channel between the TX of the XL FD node and the UE is modeled as follows:
\begin{align}
    \label{eqn:DL_chan}
    [\h_{\rm DL}]_{i_n} \triangleq \alpha_{i,n} \exp\left(\frac{\jmath2\pi}{\lambda} r_{i,n}\right),
\end{align}
where $\alpha_{i,n}$ models the attenuation factor including molecular absorption~\cite{tarboush2021teramimo}, $r_{i,n}$ represents the distance between the UE and the $n$-th antenna element of the $i$-th TX array's RF chain, and $\lambda$ is the wavelength. The UE and the passive radar target are characterized by their spherical coordinates $(r_{\rm UE},\theta_{\rm UE},\varphi_{\rm UE})$ and $(r_{\rm tar},\theta_{\rm tar},\varphi_{\rm tar})$, representing the distances from the origin as well as the elevation and azimuth angles, respectively. To this end, the distance $r_{i,n}$ included in the DL channel vector in~\eqref{eqn:DL_chan} can be computed as follows:
\begin{align}\label{eq: dist}
     r_{i,n} =&\Big((r_{\rm UE}\sin\theta_{\rm UE}\cos\varphi_{\rm UE} +\frac{d_{\rm P}}{2}+(i\!-\!1)d_{\rm RF})^2 +\\ &(r_{\rm UE}\sin\theta_{\rm UE}\sin\varphi_{\rm UE})^2 + (r\cos\theta_{\rm UE}-(n-1)d_{\rm E})^2\Big)^{\frac{1}{2}},\nonumber
\end{align}
where $d_{\rm P}$ is the distance between the TX and RX antennas, while, within a TX UPA, $d_{\rm RF}$ and $d_{\rm E}$ denote the spacing between horizontally and vertically adjacent antenna elements, respectively. Similarly holds for the case of the RX UPA.

The $N_{\rm R}\times N_{\rm T}$ end-to-end MIMO channel between the RX and TX antenna elements of the XL FD MIMO node, incorporating reflections from both the UE ($\H_{\rm UE}\in\mathbb{C}^{N_{\rm R}\times N_{\rm T}}$) and the passive radar target ($\H_{\rm tar}\in\mathbb{C}^{N_{\rm R}\times N_{\rm T}}$) (when both treated as point sources~\cite{liu2023near}), can be represented as follows:
\begin{align}\label{eq:H_R}
    \nonumber\H_{\rm R}\triangleq& \underbrace{\beta_{\rm UE} \a_{\rm RX}(r_{\rm UE},\theta_{\rm UE},\varphi_{\rm UE})\a_{\rm TX}^{\rm H}(r_{\rm UE},\theta_{\rm UE},\varphi_{\rm UE})}_{\triangleq\H_{\rm UE}}
    \\& +\underbrace{\beta_{\rm tar} \a_{\rm RX}(r_{\rm tar},\theta_{\rm tar},\varphi_{\rm tar})\a_{\rm TX}^{\rm H}(r_{\rm tar},\theta_{\rm tar},\varphi_{\rm tar})}_{\triangleq\H_{\rm tar}},
\end{align}
where $\beta_k$, with $k\triangleq\{{\rm UE},{\rm tar}\}$, is the complex-valued reflection coefficient, 
whereas the $N_{\rm T}$- and $N_{\rm R}$-element complex-valued vectors $\a_{\rm str}(\cdot)$, with ${\rm str}\triangleq\{{\rm TX},{\rm RX}\}$, are defined $\forall i=1,\ldots,N_{\rm T}^{\rm RF}(N_{\rm R}^{\rm RF})$ and $\forall n=1,\ldots,N_{\rm T}(N_{\rm R})$ as:
\begin{align}
    \label{eq:response_vec_TX}
    [\a_{\rm str}(r,\theta,\phi)]_{i_n} \triangleq a_{i,n}\exp\Big({\jmath\frac{2\pi}{\lambda}r_{i,n}}\Big).
\end{align}
In this expression, the attenuation factor $\alpha_{i,n}$ is computed similar to~\eqref{eqn:DL_chan} and the distance $r_{i,n}$ is computed analogously to \eqref{eq: dist}, adjusting for either the DL or the reflected transmission.

\subsection{Received Signal Model}
The complex-valued baseband received signal at the UE per transmitted symbol $s$ in the DL direction can be expressed as:
\begin{align}
    y \triangleq \h_{\rm DL}\W_{\rm TX}\v s + n,
\end{align}
where $n\sim\mathcal{CN}(0,\sigma^2)$ represents the Additive White Gaussian Noise (AWGN). We henceforth make the common assumption that $\h_{\rm DL}$ can be reliably estimated at the XL FD MIMO node.

Suppose that $T$ data transmissions $\s \triangleq [s(1),\ldots,s(T)]$, with $t\in\{1,\ldots, T\}$ indicating the time interval per symbol transmission, take place in the DL per coherent channel block. The baseband received signal $\Y\in\Compl^{N^{\rm RF}_{\rm R}\times T}$ after all those symbols' communication at the outputs of the RF chains of the RX of the XL FD node is mathematically expressed as:
\begin{align}\label{eq:received_matrix}
    \Y \triangleq \left(\W^{\rm H}_{\rm RX}\H_{\rm FD}\W_{\rm TX} +\D\right)\v\s+\N,
\end{align}
where $\Y = [\y(1),\ldots,\y(T)]$ and $\N \triangleq [\n(1),\ldots,\n(T)]$, with each $\n(t)\sim\mathcal{CN}(\mathbf{0}_{N^{\rm RF}_{\rm R}\times1},\sigma^2\mathbf{I}_{N^{\rm RF}_{\rm R}})$ being the $t$-th slot AWGN vector. The channel matrix $\H_{\rm FD}\in\mathbb{C}^{N_{\rm R}\times N_{\rm T}}$ is defined as $\H_{\rm FD} \triangleq \H_{\rm R} + \H_{\rm SI}$, where $\H_{\rm SI} \in \mathbb{C}^{N_{\rm R} \times N_{\rm T}}$ represents the direct SI channel between the TX and RX UPAs~\cite{alexandropoulos2017joint,FD_HMIMO_2023}, while the previously defined $\H_{\rm R}$ encapsulates the spatial parameters of the UE and the target. The digital SI canceller $\D\in\mathbb{C}^{\rm N^{\rm RF}_{\rm R} \times N^{\rm RF}_{\rm T}}$ in~\eqref{eq:received_matrix} is intended to mitigate the received SI signal, while leaving unaltered the target's parameters within $\H_{\rm R} $ for estimation. It is noted that, similar to \cite{FD_HMIMO_2023,spawc2024,gavras2024joint,islam2019unified,bayraktar2023hybrid}, we do not consider analog SI cancellation~\cite{FD_MIMO_VTM2022} requiring extra hardware components, and we will rely on the joint design of the A/D TX BF and the analog RX BF matrices for extensive numbers of TX and RX antennas, together with the digital SI cancellation matrix, targeting ISAC functionality. 
To this end, we assume that the estimates $(\widehat{r}_{\rm UE},\widehat{\theta}_{\rm UE},\widehat{\varphi}_{\rm UE})$ of the UE spatial parameters are available per coherent channel block at the XL FD MIMO node through Sounding Reference Signals (SRSs)~\cite{5G_NR_positioning}, and can be used to generate an estimate for $\H_{\rm UE}$ as $\widehat{\H}_{\rm UE}\triangleq\a_{\rm RX}(\widehat{r}_{\rm UE},\widehat{\theta}_{\rm UE},\widehat{\varphi}_{\rm UE})\a_{\rm TX}^{\rm H}(\widehat{r}_{\rm UE},\widehat{\theta}_{\rm UE},\widehat{\varphi}_{\rm UE})$. Note that the reflection coefficient $\beta_{\rm UE}$ included in $\H_{\rm UE}$ is unknown and cannot be estimated from those sounding signals. Consequently, making the common assumption that $\H_{\rm SI}$ can be reliably estimated \cite{islam2019unified,bayraktar2023hybrid,bayraktar2023self}, $\H_{\rm FD}$ appearing in~\eqref{eq:received_matrix} can be expressed as $\widehat{\H}_{\rm FD}\triangleq \widehat{\H}_{\rm UE}+\H_{\rm tar}+\H_{\rm SI}$ with $\H_{\rm tar}$ being the rank-$1$ matrix including the unknown target parameters $(r_{\rm tar},\theta_{\rm tar},\varphi_{\rm tar})$ that need to be estimated. 


\section{Proposed FD-Enabled ISAC Framework}\label{Sec: Opt}
In this section, we first derive the PEB with respect to the unknown spatial parameters of the radar target. Then, we present our optimization formulation for the tunable parameters of the considered XL FD MIMO node aiming simultaneous near-field DL data communications and monostatic-type sensing, which is efficiently solved and its complexity is analyzed for the two investigated TX/RX analog BF architectures.

\subsection{Position Error Bound Analysis}
It is evident from \eqref{eq:received_matrix}'s inspection that, for a coherent channel block involving $T$ unit-powered symbol transmissions $\forall t=1,2,\ldots,T$, yields $T^{-1}\s\s^{\rm H}=1$. Focusing on the target parameters included in the composite vector $\boldsymbol{\zeta} \triangleq [\boldsymbol{\xi},\boldsymbol{\beta}]^{\rm T}$, with $\boldsymbol{\xi}\triangleq[r_{\rm tar},\theta_{\rm tar},\phi_{\rm tar}]$ and $\boldsymbol{\beta}\triangleq[\beta_r,\beta_\iota]$, where $\beta_r \triangleq \Re\{\beta_{\rm tar}\}$ and $\beta_\iota \triangleq \Im\{\beta_{\rm tar}\}$, the received signal at the outputs of the RX RF chains of the considered XL FD node can be modeled as $\y\triangleq{\rm vec}\{\Y\}\sim\mathcal{CN}(\boldsymbol{\mu},\sigma^2)$, with mean $\boldsymbol{\mu} \triangleq {\rm vec}\{\W^{\rm H}_{\rm RX}\H_{\rm tar}\W_{\rm TX}\v\s\}$. The $5\times5$ Fisher Information Matrix (FIM)\cite{kay1993fundamentals} for the estimation of these parameters is defined~as:
\begin{align*}
    \mathbfcal{I} =\begin{bmatrix}
    \mathbfcal{I}_{r_{\rm tar}r_{\rm tar}} & \mathbfcal{I}_{r_{\rm tar}\theta_{\rm tar}} &
    \mathbfcal{I}_{r_{\rm tar}\phi_{\rm tar}} &
    \mathbfcal{I}_{r_{\rm tar}\beta_r} & \mathbfcal{I}_{r_{\rm tar}\beta_\iota}\\
    \mathbfcal{I}_{\theta_{\rm tar}r_{\rm tar}} & \mathbfcal{I}_{\theta_{\rm tar}\theta_{\rm tar}} &
    \mathbfcal{I}_{\theta_{\rm tar}\phi_{\rm tar}} &
    \mathbfcal{I}_{\theta_{\rm tar}\beta_r} & \mathbfcal{I}_{\theta_{\rm tar}\beta_\iota}\\
    \mathbfcal{I}_{\phi_{\rm tar}r_{\rm tar}} & \mathbfcal{I}_{\phi_{\rm tar}\theta_{\rm tar}} &
    \mathbfcal{I}_{\phi_{\rm tar}\phi_{\rm tar}} &
    \mathbfcal{I}_{\phi_{\rm tar}\beta_r} & \mathbfcal{I}_{\phi_{\rm tar}\beta_\iota} \\
    \mathbfcal{I}_{\beta_r r_{\rm tar}} & \mathbfcal{I}_{\beta_r\theta_{\rm tar}} &
    \mathbfcal{I}_{\beta_r\phi_{\rm tar}} &
    \mathbfcal{I}_{\beta_r\beta_r} & \mathbfcal{I}_{\beta_r\beta_\iota} \\
    \mathbfcal{I}_{\beta_\iota r_{\rm tar}} & \mathbfcal{I}_{\beta_\iota\theta_{\rm tar}} &
    \mathbfcal{I}_{\beta_\iota\phi_{\rm tar}} &
    \mathbfcal{I}_{\beta_\iota\beta_r} & \mathbfcal{I}_{\beta_\iota\beta_\iota} 
    \end{bmatrix}
\end{align*}
with its $(i,j)$-th element, where $i,j\in\boldsymbol{\zeta}$, given by: 
\begin{align}\label{eq:FIM_element}
    [\mathbfcal{I}]_{i,j}\!=\! \frac{2}{\sigma^2}\Re\left\{\!\frac{\partial \boldsymbol{\mu}^{\rm H}}{\partial[\boldsymbol{\zeta}]_i}\frac{\partial \boldsymbol{\mu}}{\partial[\boldsymbol{\zeta}]_j}\!\right\}.
\end{align}
Using the vectorized form of the received signal matrix in~\eqref{eq:received_matrix} and the target's response matrix defined in~\eqref{eq:H_R}, the partial derivatives in~\eqref{eq:FIM_element} are computed as follows:
\begin{align}\label{eq: deriv_mean}
    \frac{\partial\boldsymbol{\mu}}{\partial[\boldsymbol{\zeta}]_i} = {\rm vec}\left\{\W^{\rm H}_{\rm RX}\frac{\partial\H_{\rm tar}}{\partial[\boldsymbol{\zeta}]_i}\W_{\rm TX}\v\s\right\}.
\end{align}
To this end, for both considered analog BF architectures, the value of each $[\mathbfcal{I}]_{i,j}$ FIM element becomes:
\begin{align}
    &\nonumber[\mathbfcal{I}]_{i,j} = \\&\nonumber\frac{2T}{\sigma^2}\Re\Bigg\{{\rm Tr}\left\{\v^{\rm H}\W_{\rm TX}^{\rm H}\frac{\partial\H_{\rm tar}^{\rm H}}{\partial[\boldsymbol{\zeta}]_i}\W_{\rm RX}\W^{\rm H}_{\rm RX}\frac{\partial\H_{\rm tar}}{\partial[\boldsymbol{\zeta}]_j}\W_{\rm TX}\v\right\}\Bigg\}.
\end{align}

We now rewrite the previously derived $5\times5$ FIM expression using the submatrix definitions $\mathbfcal{I}_{\boldsymbol{\xi}\boldsymbol{\xi}} \triangleq [\mathbfcal{I}]_{1:3,1:3}$, $\mathbfcal{I}_{\boldsymbol{\xi}\boldsymbol{\beta}} \triangleq [\mathbfcal{I}]_{1:3,4:5}$, and $\mathbfcal{I}_{\boldsymbol{\beta}\boldsymbol{\beta}}\triangleq[\mathbfcal{I}]_{4:5,4:5}$ as the following compact $2\times2$ block matrix:
\begin{align}\label{eq: Block_FIM}
    \mathbfcal{I} =\begin{bmatrix}
    \mathbfcal{I}_{\boldsymbol{\xi}\boldsymbol{\xi}} & \mathbfcal{I}_{\boldsymbol{\xi}\boldsymbol{\beta}} \\
    \mathbfcal{I}_{\boldsymbol{\xi}\boldsymbol{\beta}}^{\rm T} & \mathbfcal{I}_{\boldsymbol{\beta}\boldsymbol{\beta}}
    \end{bmatrix},
\end{align}
Using the properties of the Schur's complement, the PEB for the radar target with the unknown polar coordinates in $\boldsymbol{\xi}$ is derived as follows:
\begin{align}\label{eq: PEB}
    {\rm PEB}_{{\boldsymbol{\xi}}} 
    \triangleq\sqrt{{\rm CRB}_{{\boldsymbol{\xi}}}} = \sqrt{{\rm Tr}\left\{\left[\mathbfcal{I}_{\boldsymbol{\xi}\boldsymbol{\xi}}-\mathbfcal{I}_{\boldsymbol{\xi}\boldsymbol{\beta}}\mathbfcal{I}_{\boldsymbol{\beta}\boldsymbol{\beta}}^{-1}\mathbfcal{I}_{\boldsymbol{\xi}\boldsymbol{\beta}}^{\rm T}\right]^{-1}\right\}},
\end{align}
where CRB is the Cram\'{e}r-Rao bound for the $\boldsymbol{\xi}$ estimation.

\subsection{XL FD MIMO ISAC Design Objective}
We henceforth focus on the joint design of the A/D TX BF matrix/vector, the analog RX BF matrix, and the digital SI cancellation
matrix at the considered XL FD MIMO for simultaneous maximization of the achievable DL rate and guaranteed estimation accuracy of the target's spatial parameters $\boldsymbol{\xi}$ through monostatic sensing operation, which we mathematically express via the optimization problem:
\begin{align}
        \mathcal{OP}:\nonumber&\underset{\substack{\W_{\rm TX},\W_{\rm RX}\\\D,\v}}{\max} \,\,\log_2\left(1+\sigma^{-2}\left\|\h_{\rm DL}\W_{\rm TX}\v\right\|^2\right)\\
        &\nonumber\text{\text{s}.\text{t}.}\,\left\|\W_{\rm TX}\v\right\|^2 \leq P_{\rm{\max}},\, w^{\rm TX}_{i,n}, w^{\rm RX}_{i,n} \in \mathcal{W},
        \\&\nonumber\,\quad\,\left\|\W^{\rm H}_{\rm RX}\left(\hat{\H}_{\rm UE}+\H_{\rm SI}\right)\W_{\rm TX}\v\right\|^2\leq \gamma_{\rm A},
        \\&\nonumber\,\quad\,\left\|\left(\W^{\rm H}_{\rm RX}\left(\hat{\H}_{\rm UE}+\H_{\rm SI}\right)\W_{\rm TX}+\D\right)\v\right\|^2\leq \gamma_{\rm D},
        \\&\,\quad\,\,{\rm PEB}_{{\boldsymbol{\xi}}}\leq\gamma_s.\nonumber
\end{align}
In this formulation, $\gamma_{\rm A}$ represents the threshold for the permittable instantaneous residual interference in the RX of the XL FD node after analog RX BF to avoid saturation at the respective RF chains~\cite{A2021}, $\gamma_{\rm D}$ is the respective threshold after applying digital SI cancellation~\cite{islam2019unified}, and $\gamma_{\rm s}$ indicates the level of required accuracy for the estimation of the target parameter $\boldsymbol{\xi}$ which is applied to the PEB derived in~\eqref{eq: PEB}. Finally, $\mathcal{W}$ represents any of the two codebooks $\{\mathcal{W}_{\rm PS},\mathcal{W}_{\rm L}\}$ depending on the considered analog RX BF architectures.

To avoid error accumulation in the ISAC design due to the incomplete estimation of $\H_{\rm UE}$ (recall that $\hat{\H}_{\rm UE}$ does not include $\beta_{\rm UE}$) and simplify the digital SI cancellation design, we propose to solve the following simplified version of $\mathcal{OP}$:
\begin{align}
        \mathcal{OP}_{\rm ISAC}:\nonumber&\underset{\substack{\W_{\rm TX},\W_{\rm RX},\v}}{\max} \,\,\log_2\left(1+\sigma^{-2}\left\|\h_{\rm DL}\W_{\rm TX}\v\right\|^2\right)\\
        &\nonumber\text{\text{s}.\text{t}.}\,\left\|\W_{\rm TX}\v\right\|^2 \leq P_{\rm{\max}},\, w^{\rm TX}_{i,n}, w^{\rm RX}_{i',n'} \in \mathcal{W},
        \\&\nonumber\,\quad\,\left\|\W^{\rm H}_{\rm RX}\H_{\rm SI}\W_{\rm TX}\v\right\|^2\leq \gamma_{\rm A},
        \\&\,\,\quad\,{\rm PEB}_{{\boldsymbol{\xi}}}\leq\gamma_s,\nonumber
\end{align}
which only requires the availability of $\H_{\rm SI}$ estimation. Then, the $\W_{\rm TX}$ and $\W_{\rm RX}$ solving $\mathcal{OP}_{\rm ISAC}$ can be straightforwardly used to design the digital SI cancellation matrix as $\D=-\W^{\rm H}_{\rm RX}\H_{\rm SI}\W_{\rm TX}$. It is noted, however, that the resulting XL FD MIMO ISAC design ignores the UE position estimation $(\widehat{r}_{\rm UE},\widehat{\theta}_{\rm UE},\widehat{\varphi}_{\rm UE})$. This estimation will be actually leveraged later on by our approach to estimate the target's $\boldsymbol{\xi}$-parameter.

\subsubsection{Solution for XL FD MIMO with FCPS}\label{sec: hybrid}
$\mathcal{OP}_{\rm ISAC}$ is a highly coupled and non-convex problem, thus, attaining its optimal solution is cumbersome. To deal with this challenge, we relax the problem's objective function and constraints via SemiDefinite Relaxation (SDR). To this end, 
we introduce an auxiliary variable $\J\in\Compl^{3\times3}$ with $\J\succeq0$~\cite{liu2024crb} and deploy the Schur's complement together with properties of the trace operator to reformulate $\mathcal{OP}_{\rm ISAC}$ for the FCPS case as follows:
\begin{align}
        \overline{\mathcal{OP}}_{\rm ISAC}:\nonumber&\underset{\substack{\J,\W_{\rm TX},\W_{\rm RX},\v}}{\max} \,\,{\rm Tr}\{\v^{\rm H}\W_{\rm TX}^{\rm H}\h_{\rm DL}^{\rm H}\h_{\rm DL}\W_{\rm TX}\v\}\\
        &\nonumber\text{\text{s}.\text{t}.}\,{\rm Tr}\{\v^{\rm H}\W_{\rm TX}^{\rm H}\W_{\rm TX}\v\} \leq P_{\rm{\max}},
        \\&\nonumber\,\quad\,{\rm Tr}\{\v^{\rm H}\W_{\rm TX}^{\rm H}\H_{\rm SI}^{\rm H}\W_{\rm RX}\W^{\rm H}_{\rm RX}\H_{\rm SI}\W_{\rm TX}\v\}\leq \gamma_{\rm A},
        \\&\,\quad\, \J\succeq0,\,\mathbfcal{I}_\mathbf{J}\succeq0,\,{\rm Tr}(\J^{-1})\leq\gamma_s,\nonumber
    \\&\nonumber\,\quad\,|w^{\rm TX}_{i,n}|=1, |w^{\rm RX}_{i',n'}|=1\,\,\forall i,n.
\end{align}
where $\mathbfcal{I}_\mathbf{J}$ for the FCPS case is obtained from $\mathbfcal{I}$ in~\eqref{eq: Block_FIM} as:
\begin{align}\label{eq:I_J}
    \mathbfcal{I}_\mathbf{J} \triangleq \begin{bmatrix}
    \mathbfcal{I}_{\boldsymbol{\xi}\boldsymbol{\xi}}-\J & \mathbfcal{I}_{\boldsymbol{\xi}\boldsymbol{\beta}} \\
    \mathbfcal{I}_{\boldsymbol{\xi}\boldsymbol{\beta}}^{\rm T} & \mathbfcal{I}_{\boldsymbol{\beta}\boldsymbol{\beta}}
    \end{bmatrix}.
\end{align}  
In this optimization formulation, we have rewritten Schur's complement as a positive semidefinite constraint, and the codebook constraint~\cite{liu2024crb,liu2021cramer} has been replaced with an equivalent form, according to which the magnitude of each non-zero element at the TX/RX analog BF matrix is forced to be equal to one. Next, to handle the interdependence between the A/D BF matrices, we decouple $\overline{\mathcal{OP}}_{\rm ISAC}$ excluding the codebook constraints (which will be further incorporated in the optimization) to the following subproblems which are solved until convergence via Alternating Optimization (AO):
\begin{align}
        \overline{\mathcal{OP}}_{{\rm ISAC},1}:\nonumber&\underset{\substack{\J,\F_{\rm TX}}}{\max} \,\,{\rm Tr}\{\h_{\rm DL}^{\rm H}\h_{\rm DL}\F_{\rm TX}\}\\
        &\nonumber\text{\text{s}.\text{t}.}\,{\rm Tr}\{\F_{\rm TX}\} \leq P_{\rm{\max}},\J\succeq0,
        \\&\nonumber\,\quad\,{\rm Tr}\{\H_{\rm SI}^{\rm H}\F_{\rm RX}\H_{\rm SI}\F_{\rm TX}\}\leq \gamma_{\rm A},
        \\&\,\quad\, \F_{\rm TX}\succeq0,\,\mathbfcal{I}_\mathbf{J}\succeq0,\,{\rm Tr}\{\J^{-1}\}\leq\gamma_s\nonumber,
    \\\overline{\mathcal{OP}}_{{\rm ISAC},2}:\nonumber&\underset{\substack{\J,\F_{\rm RX}}}{\min} \,\,{\rm Tr}\{\J^{-1}\}\\
        &\nonumber\text{\text{s}.\text{t}.}\,{\rm Tr}\{\H_{\rm SI}\F_{\rm TX}\H_{\rm SI}^{\rm H}\F_{\rm RX}\}\leq \gamma_{\rm A},
        \\&\,\quad\, \J\succeq0,\,\mathbfcal{I}_\mathbf{J}\succeq0,\,\F_{\rm RX}\succeq0,\nonumber
\end{align}
where we have used the matrix definitions $\F_{\rm TX}\triangleq\W_{\rm TX}\v\v^{\rm H}\W_{\rm TX}^{\rm H}$ and $\F_{\rm RX}\triangleq\W_{\rm RX}\W_{\rm RX}^{\rm H}$. Note that, according to the SDR consideration, $\F_{\rm TX}$ and $\F_{\rm RX}$ need to be rank-one and rank-$N^{\rm RF}_{\rm R}$, respectively. The latter subproblems belong to the semidefinite programming family and can be efficiently solved using standard convex problem solvers, such as CVX~\cite{cvx}. 
Interestingly, capitalizing on the findings in~\cite[Sec. 7]{luo2010semidefinite} and \cite{huang2009rank,liu2021cramer}, it can be shown that, for our optimization framework, solving the SDR subproblems is equivalent to solving their unrelaxed versions. 

When solving $\overline{\mathcal{OP}}_{{\rm ISAC},1}$ and $\overline{\mathcal{OP}}_{{\rm ISAC},2}$ iteratively, the optimal rank-one composite A/D TX BF vector $\f_{\rm opt} \in \Compl^{N_{\rm T} \times 1}$ can be obtained as $\f_{\rm opt} = \u_1\sqrt{\sigma_1}$, where $\sigma_1$ denotes the largest singular value of the unconstrained $\F_{\rm TX}$ (i.e., the codebook-based constraint for $\W_{\rm TX}$ is neglected) solving the former problem with $\u_1$ being its corresponding singular vector. To decompose $\f_{\rm opt}$ to the required A/D TX BF vectors, while respecting the unit-modulus constraint for $\W_{\rm TX}$'s elements,  we solve, similar to~\cite{zhang2022beam}, the following optimization problem:  
\begin{align}
				\underset{\substack{\W_{\rm TX},\v}}{\min} \,\,\|\f_{\rm opt}-\W_{\rm TX}\v\|_{\rm F}^2\,\,\,\,\text{\text{s}.\text{t}.}\, |w_{i,n}^{\rm TX}|=1\,\forall i,n.             
\end{align}   
It is well known that, for a given $\W_{\rm TX}$, the least-squared solution $\v=(\W_{\rm TX}^{\rm H}\W_{\rm TX})^{-1}\W_{\rm TX}^{\rm H}\f_{\rm opt}$ is obtained. To derive $\W_{\rm TX}$ given the previous $\v$ solution, we set $\w\triangleq{\rm vec}(\W_{\rm TX})$ and focus on solving the optimization problem:
\begin{align}\label{eq:w_solution}
				\underset{\substack{\w}\in\mathcal{A}}{\min} \,\,\left\|\f_{\rm opt}-(\v\otimes\I_{N_{\rm T}})\w\right\|_{\rm F}^2
\end{align}
via the Riemannian conjugate gradient algorithm~\cite{sato2022riemannian,10143983}, which converges to the desired solution in the Frobenious norm sense. In~\eqref{eq:w_solution}, $\mathcal{A}\triangleq\{w_{i,n}^{\rm TX}\big||w_{i,n}^{\rm TX}|=1\, \forall i,n\}$, which is a submanifold of $\Compl^{N_{\rm T}}$, represents $\w$'s search space.

According to~\cite{sato2022riemannian}, at each $t$-th algorithmic iteration, the relationship between the current point $\w_t$ solving~\eqref{eq:w_solution} and the next point $\w_{t+1}$ is given by the expression:
\begin{align}\label{eq: next_point}
    \w_{t+1} = f(\w_t+\alpha_tp_t),
\end{align}
where $\alpha_t$ and $\p_t$ are the Armijo step size and the search direction, respectively, while $f(\cdot)$ represents the retraction operation that projects each element of vector $\w_t+\alpha_t\p_t$ to the search space $\mathcal{A}$, i.e.: $f([\w_t+\alpha_t\p_t]_i)=[\w_t+\alpha_t\p_t]_i/|[\w_t+\alpha_t\p_t]_i|$ $\forall i=1,2,\ldots,N_{\rm T}$. The search direction $\p_t$ is given by:
\begin{align}\label{eq: search_dire}
    \p_t = -{\rm grad}f(\w_t)+b_t\mathcal{T}(\p_{t-1}),
\end{align}
where $b_t$ is the Polak-Ribiere parameter and ${\rm grad}f(\cdot)$ indicates $f(\cdot)$'s Riemannian gradient, which is computed as:
\begin{align}
    {\rm grad}f(\w_t)=\nabla f(\w_t)-\Re\{\nabla f(\w_t)\circ\w_t^*\}\circ\w_t
\end{align}
with the Euclidean gradient obtained as $\nabla f(\w_t)=2(\v^*\otimes\I_{N_{\rm T}})\left((\v^{\rm T}\otimes\I_{N_{\rm T}})\w_t-\f_{\rm opt}\right)$. In~\eqref{eq: search_dire}, $\mathcal{T}(\cdot)$ denotes the vector transport operation that maps $\p_{t-1}$ onto $\w_t$'s tangent space; for a given $\w_t$, $\mathcal{T}(\p_{t-1})=\p_{t-1}-\Re\{\p_{t-1}\circ\w_t^*\}\circ\w_t$. 

Finally, the matrix $\F_{\rm RX}$ solving $\overline{\mathcal{OP}}_{{\rm ISAC},2}$ (i.e., without the codebook-based constraint for $\W_{\rm RX}$ incorporated) at each AO iteration can be used to design the analog RX BF matrix as $\W_{\rm RX}=\exp\left(\jmath\angle\U_{:,1:N^{\rm RF}_{\rm R}}\right)$, where $\U\in\Compl^{N_{\rm R}\times N_{\rm R}}$ includes $\F_{\rm RX}$'s left singular vectors. It is noted that the latter operation inside the $\exp(\cdot)$ function extracts the phase components of the singular vectors of $\F_{\rm RX}$ corresponding to its $N_{\rm R}^{\rm RF}$ largest singular values. 

The overall algorithm for the proposed XL FD MIMO ISAC design with FCPS is summarized in Algorithm~\ref{alg:the_opt1}. As indicated, the devised iterative procedure necessitates the initialization of the A/D TX BF matrix/vector and analog RX BF matrix, either as random matrices/vector or using relevant BF solutions (e.g., \cite{spawc2024} or \cite{FD_HMIMO_2023}). In addition, to solve $\overline{\mathcal{OP}}_{{\rm ISAC},1}$ and $\overline{\mathcal{OP}}_{{\rm ISAC},2}$ iteratively, the PEB constraint is formulated using a priori knowledge for the target's parameter vector $\boldsymbol{\xi}$. For example, an initial estimate ${\widehat{\boldsymbol{\xi}}}$ for this vector can be obtained from the reflected DL signal during the first $T$ DL data transmissions at the beginning of the XL FD system's operation. Finally, the parameter $\epsilon\ll1$ is used for the overall convergence test of the algorithm, whereas the parameters $I_{\max}$, $J_{\max}$, and $N_{\max}$ indicate the maximum iteration steps in the internal loops.  


\begin{algorithm}[!t]
    \caption{XL FD MIMO ISAC Design with FCPS}
    \label{alg:the_opt1}
    \begin{algorithmic}[1]
        \renewcommand{\algorithmicrequire}{\textbf{Input:}}
        \renewcommand{\algorithmicensure}{\textbf{Output:}}
        \REQUIRE ${\H}_{\rm SI}$, $I_{\max}$, $J_{\max}$, $N_{\max}$, $\epsilon$, initial estimate ${\widehat{\boldsymbol{\xi}}}$, and \\
        \hspace{0.55cm}initializations for
        $\W_{{\rm TX},1}$, $\W_{{\rm RX},1}$, and $\v_1$.
        \ENSURE $\W_{\rm TX}$, $\W_{\rm RX}$, $\D$, and $\v$.
        \FOR{$i=2,\ldots,I_{\max}$}
        \STATE Solve $\overline{\mathcal{OP}}_{{\rm ISAC},1}$ to find $\f_{\rm opt}$.\\ 
        \STATE Set $\G_1 = \W_{{\rm TX},i-1}$.
            \FOR{$j=2,\ldots,J_{\max}$}
                \STATE Set $\g_{j}=(\G_{j-1}^{\rm H}\G_{j-1})^{-1}\G_{j-1}^{\rm H}\f_{\rm opt}$.\\
                \STATE Set $\w_1={\rm vec}(\G_{j-1})$ and $\p_1=-{\rm grad}f(\w_1)$.
                \FOR{$n=1,\ldots,N_{\max}-1$}
                    \STATE Compute $\w_{n+1}$ via \eqref{eq: next_point}.\\
                    \STATE Update the search direction $\p_{n+1}$ via \eqref{eq: search_dire}.\\
                \ENDFOR
                \STATE Set $\G_{j}={\rm unvec}\left(\w_{N_{\max}}\right)$.\\
            \ENDFOR
         \STATE Set $\v_i=\g_{J_{\max}}$ and $\W_{{\rm TX},i}=\G_{J_{\max}}$.
         \STATE Solve $\overline{\mathcal{OP}}_{{\rm ISAC},2}$ to find  $\W_{{\rm RX},i}$.\\
            \IF{$\|\W_{{\rm TX},i}\v_{i}-\W_{{\rm TX},i-1}\v_{i-1}\|^2\leq\epsilon$ and $\|\W_{{\rm RX},i}-\W_{{\rm RX},i-1}\|^2_F\leq\epsilon$}
                \STATE Break.
            \ENDIF
        \ENDFOR
        \STATE Set $\D = -(\W^{\rm H}_{{\rm RX},i}\H_{\rm SI}\W_{{\rm TX},i})$, $\W_{{\rm TX}}=\W_{{\rm TX},i}$,\\
        $\W_{{\rm RX}}=\W_{{\rm RX},i}$, and $\v=\v_{i}$. 
    \end{algorithmic}
\end{algorithm}

\subsubsection{Solution for XL FD MIMO with PCM}\label{Sec: DMA}\label{Sec: DMA_ARC}
Similar to the FCPS case, $\overline{\mathcal{OP}}_{\rm ISAC}$ when expressed for the PCM case is decoupled to the following subproblems (excluding the non-convex Lorentzian codebook constraint that will be considered later on) which are iteratively solved until convergence via AO:
\begin{align}
        \widetilde{\mathcal{OP}}_{{\rm ISAC},1}:\nonumber&\underset{\substack{\J,\widetilde{\F}_{\rm TX}}}{\max} \,\,{\rm Tr}\{\P_{\rm TX}^{\rm H}\h_{\rm DL}^{\rm H}\h_{\rm DL}\P_{\rm TX}\widetilde{\F}_{\rm TX}\}\\
        &\nonumber\text{\text{s}.\text{t}.}\,{\rm Tr}\{\P_{\rm TX}^{\rm H}\P_{\rm TX}\widetilde{\F}_{\rm TX}\} \leq P_{\rm{\max}},\J\succeq0,
        \\&\nonumber\,\quad\,{\rm Tr}\{\P_{\rm TX}\H_{\rm SI}^{\rm H}\P_{\rm RX}\widetilde{\F}_{\rm RX}\P_{\rm RX}^{\rm H}\H_{\rm SI}\P_{\rm TX}\widetilde{\F}_{\rm TX}\}\!\leq\!\gamma_{\rm A},
        \\&\,\quad\, \widetilde{\F}_{\rm TX}\succeq0,\,\mathbfcal{I}_\mathbf{J}\succeq0,\,{\rm Tr}\{\J^{-1}\}\leq\gamma_s\nonumber,
    \\\widetilde{\mathcal{OP}}_{{\rm ISAC},2}:\nonumber&\underset{\substack{\J,\widetilde{\F}_{\rm RX}}}{\min} \,\,{\rm Tr}\{\J^{-1}\}\\
        &\nonumber\text{\text{s}.\text{t}.}\,{\rm Tr}\{\P_{\rm RX}^{\rm H}\H_{\rm SI}\P_{\rm TX}\widetilde{\F}_{\rm TX}\P_{\rm TX}^{\rm H}\H_{\rm SI}^{\rm H}\P_{\rm RX}\widetilde{\F}_{\rm RX}\}\!\leq\!\gamma_{\rm A},
        \\&\,\quad\, \mathbfcal{I}_\mathbf{J}\succeq0,\,\widetilde{\F}_{\rm RX}\succeq0,\,\J\succeq0\nonumber,
\end{align}
where $\widetilde{\F}_{\rm TX} \triangleq \widetilde{\W}_{\rm TX}\v\v^{\rm H}\widetilde{\W}_{\rm TX}^{\rm H}$ and $\widetilde{\F}_{\rm RX} \triangleq \widetilde{\W}_{\rm RX}\widetilde{\W}_{\rm RX}^{\rm H}$, as well as $\mathbfcal{I}_\mathbf{J}$ is given by \eqref{eq:I_J}, now, for the PCM case. First, we compute as before the optimal rank-one composite A/D TX BF vector $\widetilde{\f}_{\rm opt} = \widetilde{\u}_1\sqrt{\widetilde{\sigma}_1}\in \Compl^{N_{\rm T} \times 1}$, with $\tilde{\sigma}_1$ being the largest singular value of the unconstrained $\widetilde{\F}_{\rm TX}$ solving $\widetilde{\mathcal{OP}}_{{\rm ISAC},1}$ with $\widetilde{\u}_1$ being its corresponding singular vector. Then, we decompose $\widetilde{\f}_{\rm opt}$ to obtain the A/D TX BF vectors by solving:
\begin{align*}
\underset{\substack{\widetilde{\W}_{\rm TX},\v}}{\min} \,\,\left\|\widetilde{\f}_{\rm opt}-\widetilde{\W}_{\rm TX}\v\right\|^2_{\rm F}\,\,\text{\text{s}.\text{t}.}\, w_{i,n}^{\rm TX}\in\mathcal{W}_{\rm L}\,\forall i,n.           
\end{align*}   
For a given $\widetilde{\W}_{\rm TX}$, yields $\v=(\widetilde{\W}_{\rm TX}^{\rm H}\widetilde{\W}_{\rm TX})^{-1}\widetilde{\W}_{\rm TX}^{\rm H}\widetilde{\f}_{\rm opt}$, whereas, for a given digital BF vector $\v$ and for $\widetilde{w}_{i,n}^{\rm TX}=0.5\left(\jmath+e^{\jmath\theta_{i,n}}\right)$, the following optimization problem is solved:
\begin{align}
        \nonumber&\underset{\substack{\theta_{i,n}\,\forall i,n}}{\min} \,\,\sum_{i=1}^{N_{\rm T}^{\rm RF}}\sum_{n=1}^{N_{\rm E}}\left|[\widetilde{\f}_{\rm opt}]_{i_n}-0.5\left(\jmath+e^{\jmath\theta_{i,n}}\right)[\v]_i\right|^2.           
\end{align}  
The critical points minimizing this objective function can be obtained after equating each of its derivatives with respect to the Lorentzian-constrained phase shift values $\theta_{i,n}$'s to zero, yielding the following solutions:
\begin{align}\label{eq: DMA_theta}
    \theta_{i,n}={\rm arg}\left(2\frac{[\widetilde{\f}_{\rm opt}]_{i_n}}{[\v]_i}-\jmath\right)\,\forall i,n.
\end{align}

The direct incorporation of the Lorentzian structure of the elements of the analog BF matrix in $\widetilde{\mathcal{OP}}_{\rm ISAC,1}$ is cumbersome and would lead to hard a problem due to the existing coupling between this matrix and the digital TX BF vector. However, in $\widetilde{\mathcal{OP}}_{\rm ISAC,2}$ where only the analog RX BF matrix appears as the optimization parameter, this Lorentzian constraint can be straightforwardly incorporated. Let $\widetilde{\w}_i^{\rm RX}\triangleq[\widetilde{w}_{i,1}^{\rm RX},\widetilde{w}_{i,2}^{\rm RX},\ldots,\widetilde{w}_{i,N_{\rm E}}^{\rm RX}]^{\rm T}\in\Compl^{N_{\rm E}\times1}$ include the adjustable response weights attached to the $i$-th RF chain of the RX DMA. Clearly, $\widetilde{\F}_{\rm RX}$ is a $N_{\rm R}\times N_{\rm R}$ diagonal block matrix where each $i$-th block of size $N_{\rm E}\times N_{\rm E}$ has the structure $\widetilde{\w}_i^{\rm RX}(\widetilde{\w}_i^{\rm RX})^{\rm H}$. By using the Lorentzian constraint $\widetilde{\w}_i^{\rm RX}=0.5(\jmath\boldsymbol{1}_{N_{\rm E}}+\q_i)$, with $\q_i\triangleq[q_{i,1},\ldots,q_{i,N_{\rm E}}]^{\rm T}\in\Compl^{N_{\rm E}\times1}$ (i.e., including unconstrained analog weights in the $\widetilde{\w}_i^{\rm RX}$ expression), 
$\widetilde{\mathcal{OP}}_{\rm ISAC,2}$ can be rewritten as follows~\cite{gavriilidis2024metasurface}:
\begin{align}
        \nonumber&\underset{\substack{\J,\{\Q_i\}_{i=1}^{N_{\rm R}^{\rm RF}}}}{\min} \,\,{\rm Tr}\{\J^{-1}\}\\
        &\nonumber\text{\text{s}.\text{t}.}\,\sum_{i=1}^{N_{\rm R}^{\rm RF}}{\rm Tr}\{\L_i\Q_i\}\leq \gamma_{\rm A},\,{\rm diag}(\Q_i)=\boldsymbol{1}_{\rm N_{\rm R}+1},
        \\&\,\quad\, \mathbfcal{I}_\mathbf{J}\succeq0,\,\J\succeq0,\,\Q_i\succeq0\,\forall i\nonumber.
\end{align}
where $\L_i,\Q_i\in\Compl^{(N_{\rm E}+1)\times(N_{\rm E}+1)}$ are defined as follows:
\begin{align}
&\nonumber\L_i \triangleq \begin{bmatrix}
[\A]_{i_1:iN_{\rm E},i_1:iN_{E}}&\jmath\boldsymbol{1}_{N_{\rm E}}\\
(\jmath\boldsymbol{1}_{N_{\rm E}})^{\rm H} & 1
\end{bmatrix},\,\,\Q_i\triangleq\begin{bmatrix}
\q_i \\
1
\end{bmatrix}
\begin{bmatrix}
\q_i^{\rm H}& 1
\end{bmatrix},
\end{align}
and $\A\triangleq\P_{\rm RX}^{\rm H}\H_{\rm SI}\P_{\rm TX}\F_{\rm TX}\P_{\rm TX}^{\rm H}\H_{\rm SI}^{\rm H}\P_{\rm RX}$. To solve this problem, we have applied a series of SDRs, where each rank-one constraint corresponding to $\Q_i$ $\forall i$ was replaced with a positive semidefinite one. Additionally, each element of $\mathbfcal{I}_\mathbf{J}$ has been reformulated in accordance with the block-diagonal structure of each $\Q_i$, similar to $\L_i$. This transformation resulted in a convex problem, which was efficiently solved using standard convex programming solvers (e.g., CVX~\cite{cvx}). Finally, $\mathbf{q}_i^{\rm opt}$ $\forall i$ solving the latter optimization problem were obtained as $\mathbf{q}_i^{\rm opt} = \exp(\jmath\angle[\u_i]_{1:N_{\rm E}})$, with $\u_i$ being the principal singular vector of $\Q_i$. It is noted that, similar to the FCPS case, the SDR-relaxed subproblems $\widetilde{\mathcal{OP}}_{\rm ISAC,1}$ and $\widetilde{\mathcal{OP}}_{\rm ISAC,2}$ yield solutions that satisfy the necessary rank constraints. 

The overall algorithm for the proposed XL FD MIMO ISAC design with PCM is summarized in Algorithm~\ref{alg:the_opt2}. Its initialization and stopping criterion are similar to Algorithm~\ref{alg:the_opt1}. Note that $\P_{\rm TX}$ and $\P_{\rm RX}$ can be obtained from the characterization of the designed TX and RX DMAs, respectively.

\begin{algorithm}[!t]
    \caption{XL FD MIMO ISAC Design with PCM}
    \label{alg:the_opt2}
    \begin{algorithmic}[1]
        \renewcommand{\algorithmicrequire}{\textbf{Input:}}
        \renewcommand{\algorithmicensure}{\textbf{Output:}}
        \REQUIRE ${\H}_{\rm SI}$, $\P_{\rm TX}$, $\P_{\rm RX}$, $I_{\rm max}$, $J_{\rm max}$, $\epsilon$, initial estimate ${\widehat{\boldsymbol{\xi}}}$, \\ \hspace{0.55cm}and initializations $\widetilde{\W}_{{\rm TX},1}$, $\widetilde{\W}_{{\rm RX},1}$, and $\v_1$ 
        \ENSURE $\widetilde{\W}_{\rm TX}$, $\widetilde{\W}_{\rm RX}$, $\D$, and $\v$.
            \FOR{$i=2,\ldots,I_{\max}$}
            \STATE Solve $\widetilde{\mathcal{OP}}_{\rm ISAC,1}$ to find $\widetilde{\f}_{\rm opt}$.
            \STATE Set $\G_1 = \widetilde{\W}_{{\rm TX},i-1}$.
            \FOR{$j=2,\ldots,J_{\max}$}
                \STATE Set $\g_{j}=(\G_{j-1}^{\rm H}\G_{j-1})^{-1}\G_{j-1}^{\rm H}\widetilde{\f}_{\rm opt}$.\\
                \STATE Construct each element of $\G_{j}$ using \eqref{eq: DMA_theta}.\\ 
            \ENDFOR
            \STATE Set $\v_i=\g_{J_{\max}}$ and $\widetilde{\W}_{{\rm TX},i}=\G_{J_{\rm max}}$.
            \STATE Solve $\widetilde{\mathcal{OP}}_{\rm ISAC,2}$ to find $\widetilde{\W}_{{\rm RX},i}$.
            \IF{$\|\widetilde{\W}_{{\rm TX},i}\v_{i}-\widetilde{\W}_{{\rm TX},i-1}\v_{i-1}\|^2\leq\epsilon$ and $\|\widetilde{\W}_{{\rm RX},i}-\widetilde{\W}_{{\rm RX},i-1}\|^2\leq\epsilon$}
                \STATE Break.
            \ENDIF
        \ENDFOR
        \STATE Set $\D = -(\widetilde{\W}^{\rm H}_{{\rm RX},i}\P^{\rm H}_{\rm RX}\H_{\rm SI}\P_{\rm TX}\widetilde{\W}_{{\rm TX},i})$,\\ $\widetilde{\W}_{{\rm TX}}=\widetilde{\W}_{{\rm TX},i}$, $\widetilde{\W}_{{\rm RX}}=\widetilde{\W}_{{\rm RX},i}$, and $\v=\v_{i}$.
    \end{algorithmic}
\end{algorithm}

\begin{figure*}[t!]
    \centering
    \begin{subfigure}[t]{0.5\textwidth}
        \centering
        \includegraphics[scale=0.55]{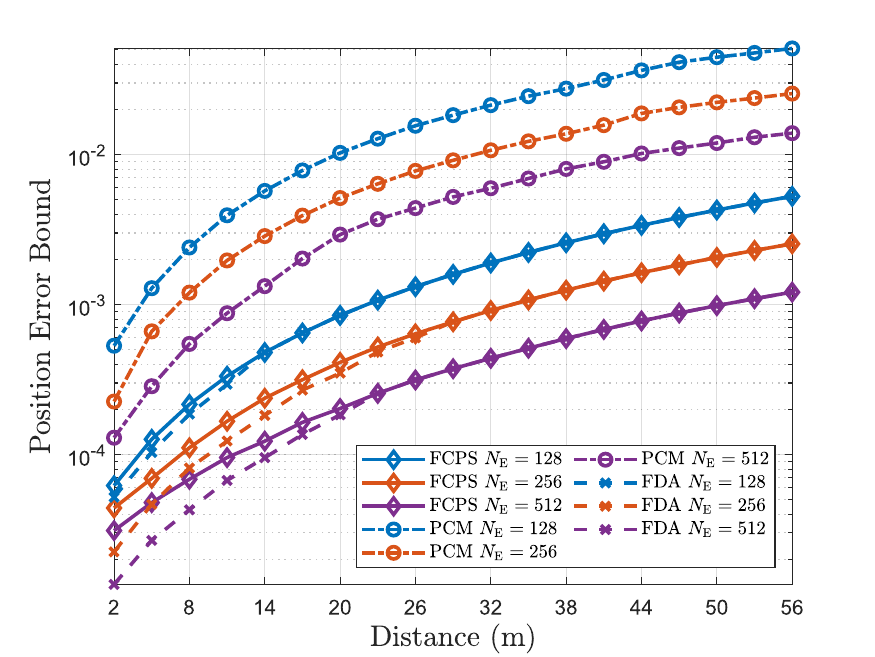}
        \caption{Position Estimation Performance.}
        \label{fig: PEB_LIM}
    \end{subfigure}%
    \begin{subfigure}[t]{0.5\textwidth}
        \centering
        \includegraphics[scale=0.55]{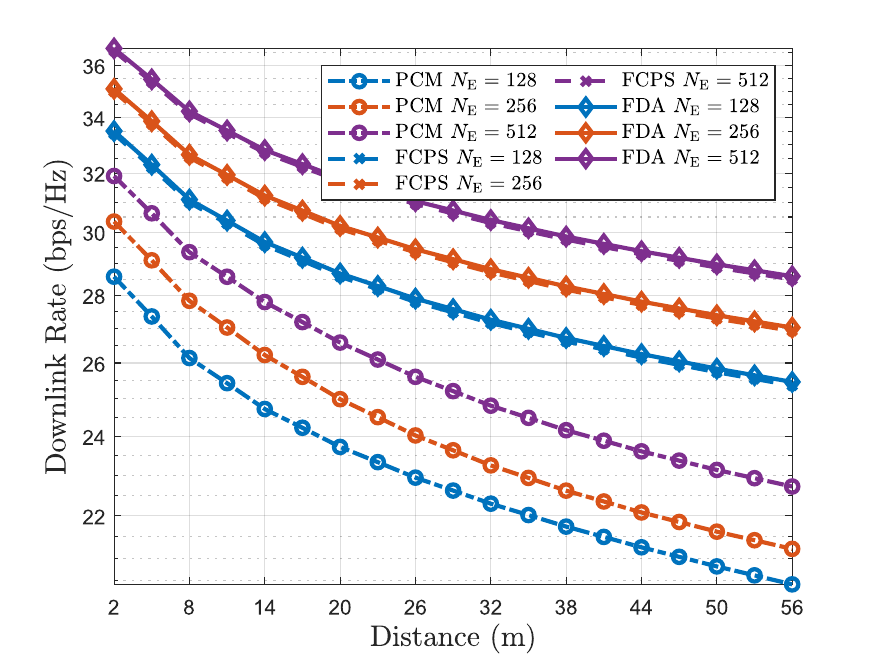}
        \caption{Communication Performance.}
        \label{fig: DL_LIM}
    \end{subfigure}
    \caption{\small{Position estimation and achievable rate performances versus the target/UE distance $r$ when the target/UE is placed at the unknown azimuth and elevation angles $\theta=30^{\circ}$ and $\phi=60^{\circ}$, respectively, considering an RX/TX UPA with $N_{\rm T}^{\rm RF}=N_{\rm R}^{\rm RF}=2$ TX/RX RF chains and $N_{\rm E}=\{128,256,512\}$ antenna elements, and $P_{\rm{\max}}=24$ dBm. The performance with RX and TX FDAs equipped with $N_{\rm R}^{\rm RF}N_{\rm E}$ RX and $N_{\rm T}^{\rm RF}N_{\rm E}$ TX RF chains, respectively, has been also evaluated.}}
\end{figure*}

\subsection{Complexity Analysis}

The convergence rate of both presented iterative Algorithms~\ref{alg:the_opt1} (FCPS case) and~\ref{alg:the_opt2} (PCM case) for XL FD MIMO ISAC depends on the numbers of TX/RX RF chains and the numbers of antenna/metamaterial elements. In both algorithms, interior-point optimization methods have deployed for the the SDR parts. It is well known that these methods exhibit a worst-case complexity of $\mathcal{O}(n^2\sum_{i=1}^{C}m_i^2+n\sum_{i=1}^{C}m_i^3)$~\cite{nemirovski2004interior,boyd2004convex}, where $C$ is the number of Linear Matrix Inequalities (LMIs), $m_i$ represents the dimension of the $i$-th LMI constraint, and $n$ is the number of optimization variables. In the context of Algorithm~\ref{alg:the_opt1} dealing with $\overline{\mathcal{OP}}_{\rm ISAC,1}$, we have that $n=N_{\rm T}^2+9$, $C=3$ (in particular, due to the constraints $\J\succeq 0$, $\boldsymbol{\mathcal{I}}_{\J}\succeq 0$, and $\F_{\rm TX}\succeq 0$), as well as $m_1=3$, $m_2=5$ and $m_3=N_{\rm T}$, yielding the approximate complexity of $\mathcal{O}\left(N_{\rm T}^6\right)$. Moreover, the Riemannian conjugate gradient algorithm requires a complexity of $\mathcal{O}\left(N_{\max}(N_{\rm T}N^{\rm RF}_{\rm T})^{1.5}\right)$ \cite{10143983}, while the least-squared solution has a complexity of $\mathcal{O}\left(N_{\rm T}\left(N_{\rm T}^{\rm RF}\right)^2\right)$. Following the same reasoning for $\overline{\mathcal{OP}}_{\rm ISAC,2}$, its approximate complexity is $\mathcal{O}\left(N_{\rm R}^6\right)$. Taking all above into account, the overall worst-case complexity of Algorithm~\ref{alg:the_opt1} is given by:
\begin{align}
    \nonumber&\mathcal{O}\Big(I_{\max}\Big(\left(N_{\rm T}^6+N_{\rm R}^6\right)\\
    \nonumber&\quad\,+J_{\max}\left(N_{\rm T}\left(N_{\rm T}^{\rm RF}\right)^2+N_{\max}\left(N_{\rm T}N_{\rm T}^{\rm RF}\right)^{1.5}\right)\Big)\Big).
\end{align}

Similarly, in Algorithm~\ref{alg:the_opt2}, $\widetilde{\mathcal{OP}}_{\rm ISAC,1}$ and $\widetilde{\mathcal{OP}}_{\rm ISAC,2}$ exhibit the approximate complexities of $\mathcal{O}\left(N_{\rm T}^6\right)$ and $\mathcal{O}\left(\left(N_{\rm R}^{\rm RF}\right)^3N_E^6\right)$, respectively. The least-squared solution has the same complexity as the FCPS case, while constructing the analog Lorentzian-constrained weights yields a complexity of $\mathcal{O}\left(N_{\rm T}\right)$. Therefore, the total worst-case complexity of Algorithm 2 is evaluated as follows:
\begin{align}
    &\nonumber\mathcal{O}\Big(I_{\max}\Big(\Big(N_{\rm T}^6+\left(N_{\rm R}^{\rm RF}\right)^3N_E^6\Big)\\
    \nonumber&\quad\,+J_{\max}\left(N_{\rm T}\left(N_{\rm T}^{\rm RF}\right)^2+N_{\rm T}\right)\Big)\Big).
\end{align}

\section{Numerical Results and Discussion}\label{Sec: Numerical}
In this section, we present extensive simulation results for the performance of the proposed FD-enabled framework for simultaneous DL data communications and monostatic-type sensing, considering various system parameter settings. The trade-off among communications, sensing, and SI mitigation is numerically assessed for different scenarios.

\subsection{Simulation Parameters}
The proposed XL FD MIMO node was assumed to operate at the central frequency $120$~GHz with a narrow bandwidth of $B=150$~kHz. The distance between its adjacent horizontal and vertical antenna elements was set to $d_{\rm RF} =\lambda/2$ and $ d_{\rm E} = \lambda/5$, respectively, and the separation between its TX and RX panels was fixed at $d_{\rm P} = 0.5$ meters. We have assumed that a coherent channel block spans $T = 200$ transmissions, and that both the UE and the target are randomly positioned at a fixed elevation angle $\theta = 30^\circ$, an azimuth angle $\phi \in [0^\circ, 180^\circ]$, and a range $r\in [2, 50]$ in meters.
For the actual estimation of the target spatial parameters with our ISAC-optimized XL FD MIMO ISAC design, we have used MUltiple SIgnal Classification (MUSIC)~\cite{millar2011maximum} estimation. As mentioned earlier, this MUSIC estimation was combined with the availability of the estimation $(\widehat{r}_{\rm UE},\widehat{\theta}_{\rm UE},\widehat{\varphi}_{\rm UE})$ to distinguish the target position estimation with that of the UE. All simulation results were obtained using $500$ Monte Carlo runs. The AWGN's variance was set as $\sigma^2=-84$ dBm, and $\beta_k$ $\forall k\in\{\rm UE,tar\}$ in \eqref{eq:H_R} was chosen randomly with unit amplitude. In Algorithms 1 and 2, we assumed a convergence accuracy level of $\epsilon=10^{-2}$. For comparisons, we have also simulated the performance of the recent ISAC designs in~\cite{spawc2024} and~\cite{FD_HMIMO_2023} and considered all compared frameworks with both analog BF architectures, as well as the FDA array architecture at the TX and RX of the XL FD MIMO node.

\begin{figure*}[t!]
    \centering
    \begin{subfigure}[t]{0.5\textwidth}
        \centering
        \includegraphics[scale=0.55]{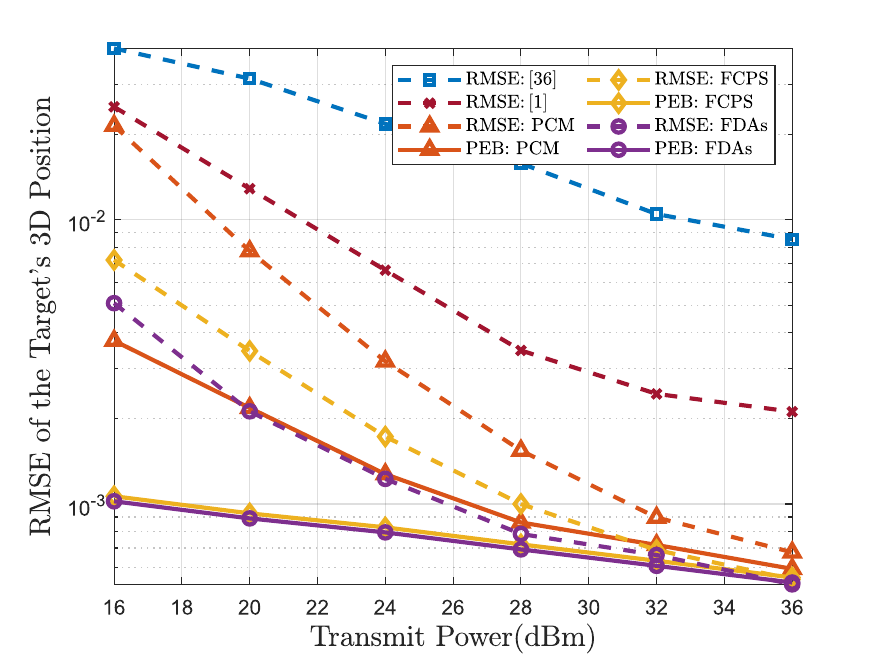}
        \caption{RMSE of Position Estimation.}
        \label{fig: RMSE}
    \end{subfigure}%
    \begin{subfigure}[t]{0.5\textwidth}
        \centering
        \includegraphics[scale=0.55]{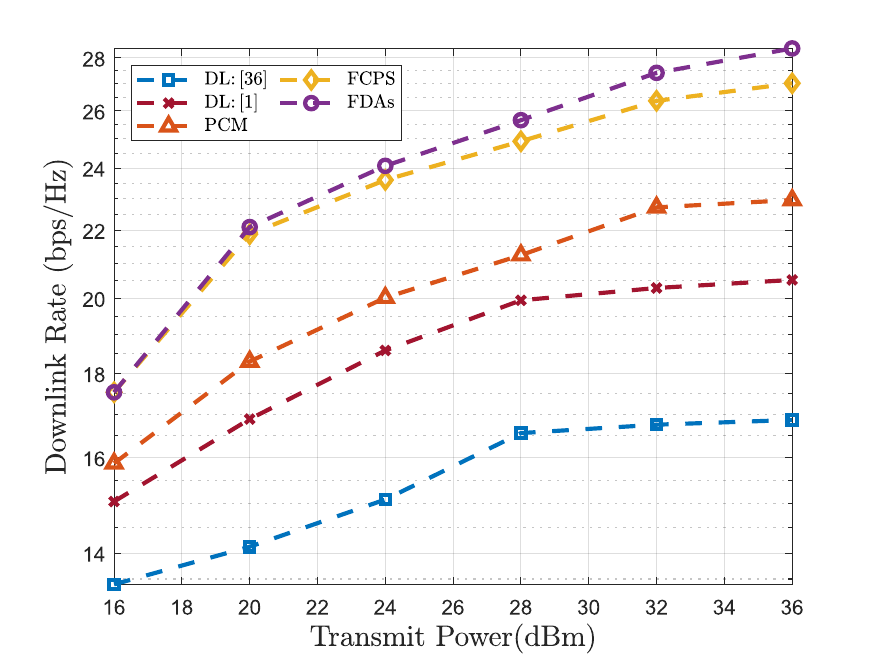}
        \caption{Achievable DL Rate.}
        \label{fig: DL_rate}
    \end{subfigure}
    \caption{\small{Position estimation and achievable DL rate performances versus $P_{\rm{\max}}$ in dBm for the proposed XL FD MIMO ISAC designs with FCPS and PCM analog BF architectures, considering $N_{\rm T}^{\rm RF}=N_{\rm R}^{\rm RF}=2$ TX/RX RF chains, $N_{\rm E}=128$ antenna elements, $\gamma_{\rm A}=10^{-6}$, and $\gamma_s=10^{-3}$. The respective performances with RX and TX FDAs equipped with $128$ RX and $128$ TX RF chains, respectively, have been also evaluated, as well as those with the ISAC schemes in~\cite{FD_HMIMO_2023} and~\cite{spawc2024}.}}
    \label{fig: RMSE_DL_rate}
\end{figure*}

\begin{figure}
        \includegraphics[scale=0.6]{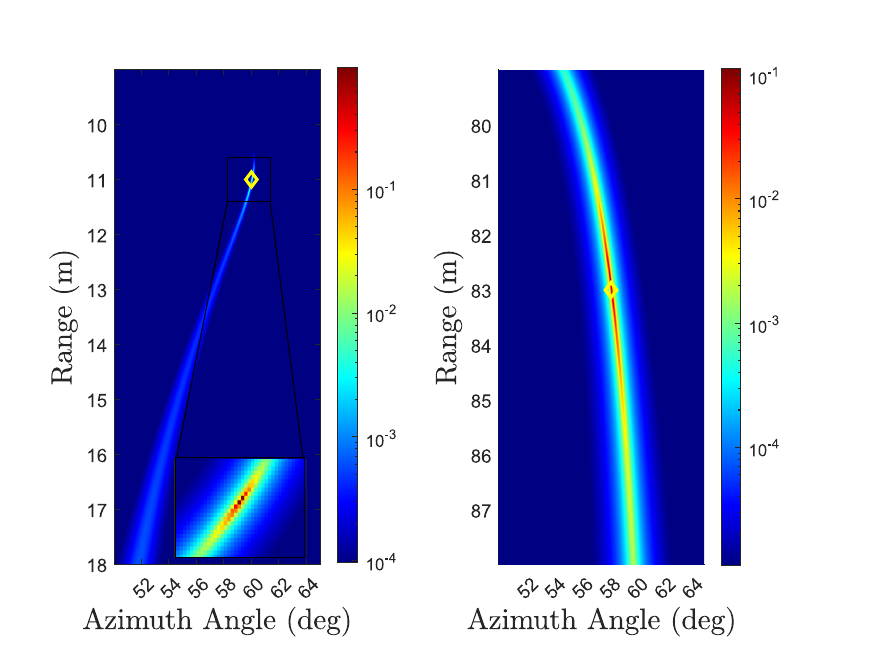}
        \caption{\small{Beamforming capability of the proposed XL FD MIMO ISAC design with the PCM analog combining architecture using the setting of parameters in Fig.~\ref{fig: RMSE_DL_rate} and $P_{\rm{\max}}=24$ dBm. The left plot shows the beam focusing achieved by the proposed scheme when the target (its actual position is indicated by the yellow diamond) lies in the FD node's near-field region, whereas the right plot depicts the beam steering accomplished for the case where UE is within the far field.}}
        \label{fig: Beam}
\end{figure}

\subsection{Communications and Sensing Performance}
To assess the boundary objectives of our ISAC framework, we have optimized the proposed XL FD MIMO system separately for PEB minimization and achievable DL rate maximization, and the respective performance results as functions of the target/UE distance $r$ in meters are illustrated in Figs.~\ref{fig: PEB_LIM} and~\ref{fig: DL_LIM}. The target/UE was placed at the unknown azimuth and elevation angles $\theta=30^{\circ}$ and $\phi=60^{\circ}$, respectively, the RX/TX UPA was equipped with $N_{\rm T}^{\rm RF}=N_{\rm R}^{\rm RF}=2$ RX/TX RF chains and $N_{\rm E}=\{128,256,512\}$ antenna elements, and $P_{\rm{\max}}$ was set to $24$ dBm. Both considered TX/RX analog BF architectures were simulated as well as the FDA TX/RX architecture. It can be observed from both figures that, as expected, both performance metrics degrade with increasing $r$. In addition, it is evident that the designs with FDAs and fully-connected PS networks (abbreviated as FCPS) outperform the respective ones with partially-connected arrays of metamaterials (abbreviated as PCM). This is attributed to the fact that, in general, FDA provides the maximum design flexibility, then follows FCPS, and PCM offers the least flexibility. Interestingly, for this figure's parameters, FCPS yields similar PEB and rate performances with FDA, while with PCM both of these metrics are degraded. However, this performance gap can be compensated by increasing $N_{\rm E}$ in the PCM architecture. Note that this UPA size scaling is much more cost and power efficient with PCM than FCPS and FDA. 


\begin{figure*}[t!]
    \centering
    \begin{subfigure}[t]{0.5\textwidth}
        \centering
        \includegraphics[scale=0.55]{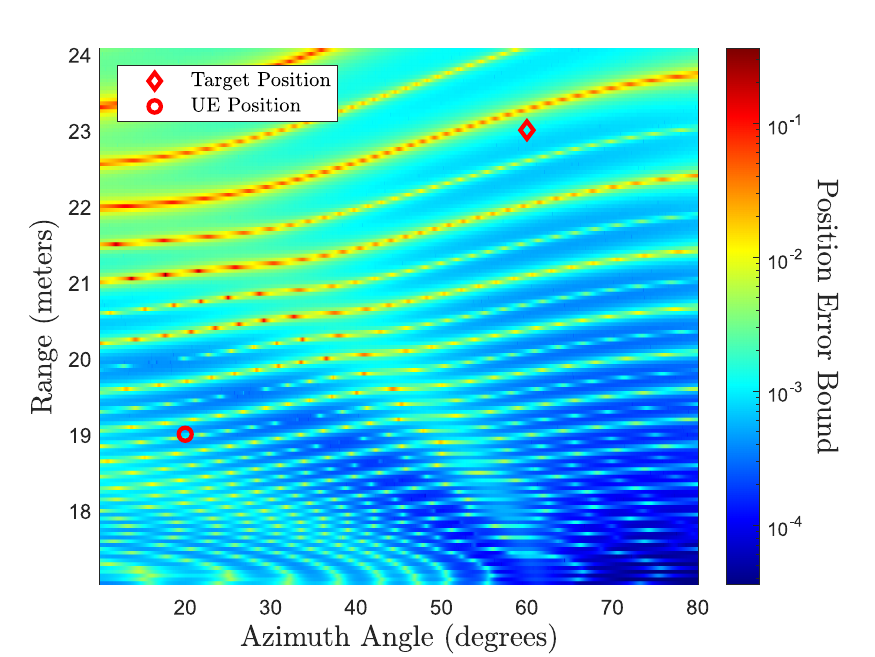}
        \caption{The FCPS case.}
        \label{fig: PEB_MAP_FCPS}
    \end{subfigure}%
    \begin{subfigure}[t]{0.5\textwidth}
        \centering
        \includegraphics[scale=0.55]{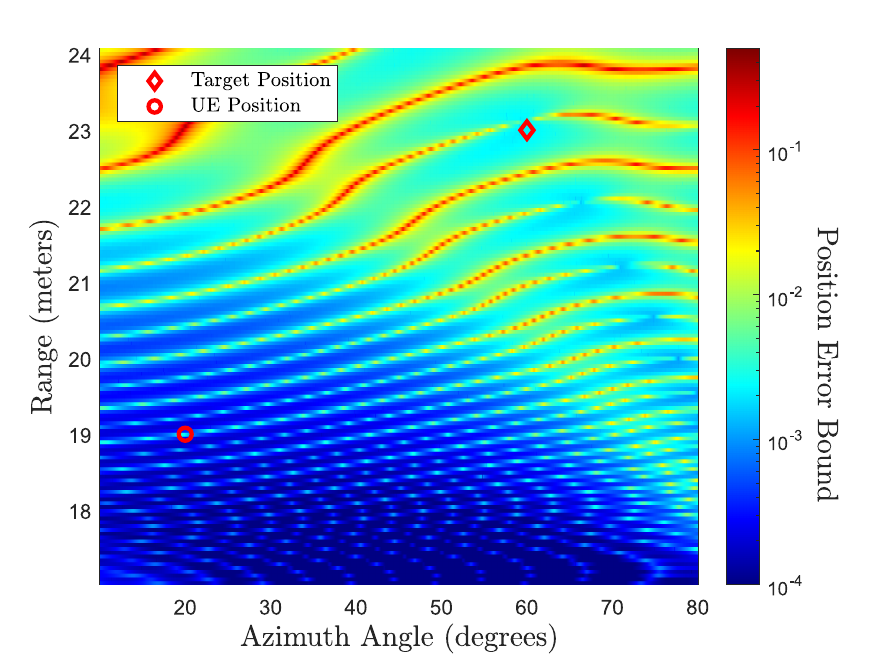}
        \caption{The PCM case.}
        \label{fig: PEB_MAP_PCM}
    \end{subfigure}
    \caption{\small{PEB performance with the proposed XL FD MIMO ISAC designs for a two-dimensional area including the true target position and that of the UE, using the setting of parameters in Fig.~\ref{fig: RMSE_DL_rate} and $P_{\rm{\max}}=24$ dBm.}}
    \label{fig: PEB_MAP}
\end{figure*}

\begin{figure}
        \includegraphics[scale=0.55]{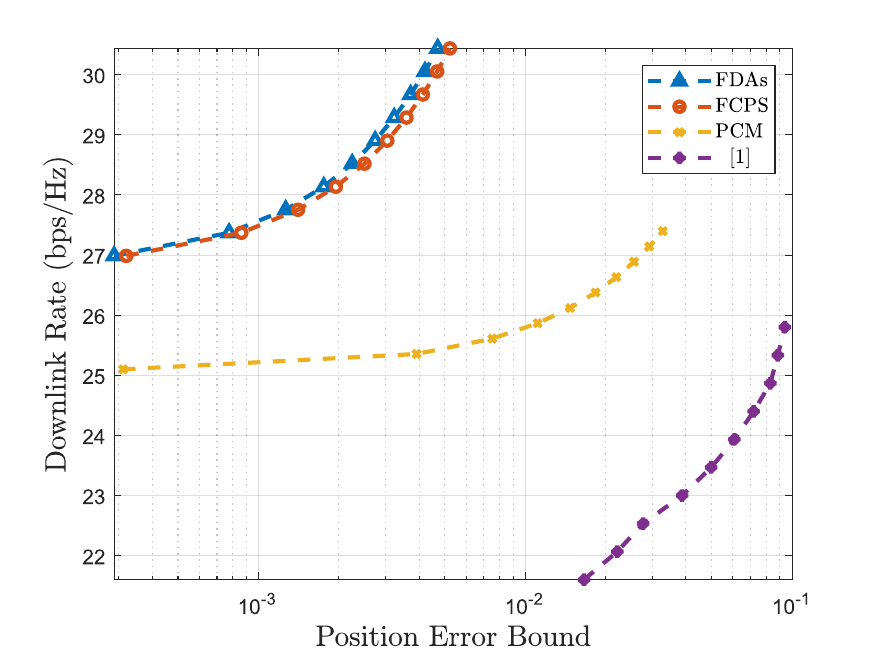}
        \caption{\small{Trade-off between the communication and sensing functionalities for the proposed XL FD MIMO system with $N_{\rm T}^{\rm RF}=N_{\rm R}^{\rm RF}=2$ TX/RX RF chains, each connected to $N_{\rm E}=128$ antenna elements, considering $\gamma_{\rm A}=10^{-6}$ and $P_{\rm{\max}}=36$ dBm. The results were obtained from $\mathcal{OP}_{\rm Pareto}$'s solution for various $\nu$ values within $[0,1]$.}}
        \label{fig: tradeoff}
\end{figure}

The Root Mean Squared Error (RMSE) for the estimation of the target's 3D position and the respective PEB as well as the achievable DL rate with the proposed ISAC scheme, considering either the FCPS (Algorithm~1) or the PCM (Algorithm~2) analog BF architecture, are depicted in Figs.~\ref{fig: RMSE} and~\ref{fig: DL_rate}, respectively, for different values of $P_{\rm{\max}}$ in dBm. It has been considered that the proposed XL FD MIMO node has $N_{\rm T}^{\rm RF}=N_{\rm R}^{\rm RF}=2$ RX/TX RF chains each with $N_{\rm E}=128$ antenna elements, the threshold for the permittable instantaneous residual interference was set as $\gamma_{\rm A}=10^{-6}$, and the PEB requirement was $\gamma_s=10^{-3}$. As observed, the proposed scheme outperforms the state-of-the-art designs~\cite{FD_HMIMO_2023} and~\cite{spawc2024} for both considered analog BF architectures at any $P_{\rm{\max}}$ level. It is shown that, as the SNR increases, both proposed designs yield improved positioning and rate performances. Interestingly, it is evident that, in this regime, the estimation's RMSE converges to the respective PEB. In addition, the effectiveness of the proposed ISAC designs is corroborated by their consistent ability to meet the PEB constraints as SNR values increase. It can be also seen that, in comparison to FDA and FCPS, the proposed design with the PCM analog BF architecture hardly meets the PEB constraint in low SNRs. This is attributed to the fact that, for the same number of antenna elements, first FDA and then FCPS provide more directive beams than PCM. As in Figs.~\ref{fig: PEB_LIM} and~\ref{fig: DL_LIM}, this can be compensated for PCM by just increasing $N_{\rm E}$ per RX/TX RF chain. Additionally, it is demonstrated that, as the number of antenna elements grows and the transmit power increases, the benchmark schemes exhibit a declining rate of improvement in terms of PEB and DL rate. This occurs because with the benchmark schemes for XL array cases, and especially at high transmit powers, the deployed SI cancellation techniques become insufficient to suppress the increasing SI signal. As a result, both the analog and digital BF must allocate more resources to SI cancellation in order to meet the respective constraints. In contrast, the proposed schemes maintain a consistent improvement in both PEB and DL rate, even with a high transmit power. This is attributed to the fact that, in the proposed optimization framework, we explicitly incorporate an SI constraint, and both the analog and digital BF components contribute jointly to SI mitigation.

In Fig.~\ref{fig: Beam}, we focus on the investigation of the BF capability of the proposed XL FD MIMO ISAC design with the PCM analog combining architecture for the sensing functionality, considering the same parameters with Fig.~\ref{fig: RMSE_DL_rate}, $P_{\rm{\max}}=24$ dBm, and two cases for the actual target position: one lying on the near-field region of the FD node (left plot) and the other on the far field (right plot). The normalized beam focusing/steering gain is actually depicted. It can be observed from the left subfigure that beam focusing is achieved around the actual target position in the near field, whereas beam steering is realized towards the actual target position in the far field. Recall that the proposed design's goal with respect to the sensing functionality focuses on minimizing the PEB under a predefined threshold. This behavior indicates that, as expected, when we traverse out of the near-field region, the beamdepth exceeds the beamwidth, thus, beam focusing transforms to beam steering. From a sensing-perspective, this implies that range estimation becomes increasingly harder as the level of range ambiguity increases. This transition in the BF behavior actually takes place within one tenth of the near-field region~\cite{liu2023near}. It is noted that inside the Rayleigh zone (left plot), the XL array must generate a spherical wavefront (i.e., a 3D beam pattern spanning over elevation, azimuth, and range) so that radiation from every antenna/metamaterial element arrives in phase at the focal point. When this 3D Fresnel ring structure is projected onto just two dimensions (i.e., the azimuth-range plane in our case here), it appears as a curved ridge, which represents the intersection of the spherical wavefront with the 2D observation plane. Once the observation point moves beyond the Rayleigh distance (right plot), the path differences flatten, the steering vector depends only on the angle, and the array response collapses to a straight vertical stripe, similar to conventional BF in the far‑field region. Thus, the curvature in the left plot of Fig.~\ref{fig: Beam} showcases the near‑field range‑angle coupling, whereas the straight lobe in Fig.~\ref{fig: Beam}'s right plot confirms the far‑field decoupling.

In Fig.~\ref{fig: PEB_MAP}, we focus on the system setup in Fig.~\ref{fig: RMSE_DL_rate} and investigate the sensing performance of our XL FD MIMO ISAC designs in a wide two-dimensional area including the true target position and that of the UE. We have specifically evaluated the PEB for the target with both our FCPS- and PCM-based designs from Fig.~\ref{fig: RMSE_DL_rate} in a single Monte Carlo iteration.
It can be observed that the PEB performance in the area surrounding the true position of the UE/target is satisfactory  for both analog BF cases. In particular, the PEB constraint of $\gamma_s = 10^{-3}$ is achieved in $80\%$ of the considered area with our FCPS-based design, whereas, for the PCM case, it is achieved in $50\%$ of the area. As in the previous numerical investigations, this difference is attributed to the lesser BF flexibility of the PCM architecture compared to the FCPS one. It can be also seen that, as expected, the PEB performance metric degrades as we move further from the XL FD MIMO node. Despite this fact, it is shown that both designs achieve adequate PEB performance even for subareas within the far-field region of the proposed XL FD MIMO node.

\subsection{Trade-Off Between Communications and Sensing}
We now focus on the evaluation of the trade-off between the communication and sensing functionalities offered by the proposed XL FD MIMO framework with respect to various system configurations. To this end, we use the definition for the achievable DL rate $\mathcal{R}\triangleq\log_2(1+\sigma^{-2}\left\|\h_{\rm DL}\W_{\rm TX}\v\right\|^2)$ and the weighting parameter $\nu\in[0,1]$ to formulate the optimization:
\begin{align}
        \mathcal{OP}_{\rm Pareto}:\nonumber&\underset{\substack{\W_{\rm TX},\W_{\rm RX},\v}}{\max} \quad \nu \frac{\mathcal{R}-\mathcal{R}_{\min}}{\mathcal{R}_{\max}-\mathcal{R}_{\min}}\\&\nonumber\qquad\qquad\quad-(1-\nu)\frac{{\rm PEB}_{\boldsymbol{\xi}}-{\rm PEB}_{\boldsymbol{\xi},\min}}{{\rm PEB}_{\boldsymbol{\xi},\max}-{\rm PEB}_{\boldsymbol{\xi},\min}}\\
        &\nonumber\text{\text{s}.\text{t}.}\,\left\|\W_{\rm TX}\v\right\|^2 \leq P_{\rm{\max}},\, w^{\rm TX}_{i,n}, w^{\rm RX}_{i,n} \in \mathcal{W},
        \\&\nonumber\,\quad\,\left\|\W^{\rm H}_{\rm RX}\H_{\rm SI}\W_{\rm TX}\v\right\|^2\leq \gamma_{\rm A},
\end{align}
where $\mathcal{R}_{\max}$, $\mathcal{R}_{\min}$ and ${\rm PEB}_{\boldsymbol{\zeta},\max}$, ${\rm PEB}_{\boldsymbol{\zeta},\min}$ represent the maximum and minimum achievable rates, as well as the maximum and minimum PEB, respectively, obtained by maximizing and minimizing $\mathcal{OP}_{\rm ISAC}$, with respect to each objective, while ignoring the constraint related to $\gamma_s$. Note that $\mathcal{OP}_{\rm Pareto}$ can be solved for the considered FCPS and PCM analog BF architectures for our XL FD MIMO node using Sections~\ref{sec: hybrid} and~\ref{Sec: DMA}, respectively.

\begin{figure}
        \includegraphics[scale=0.55]{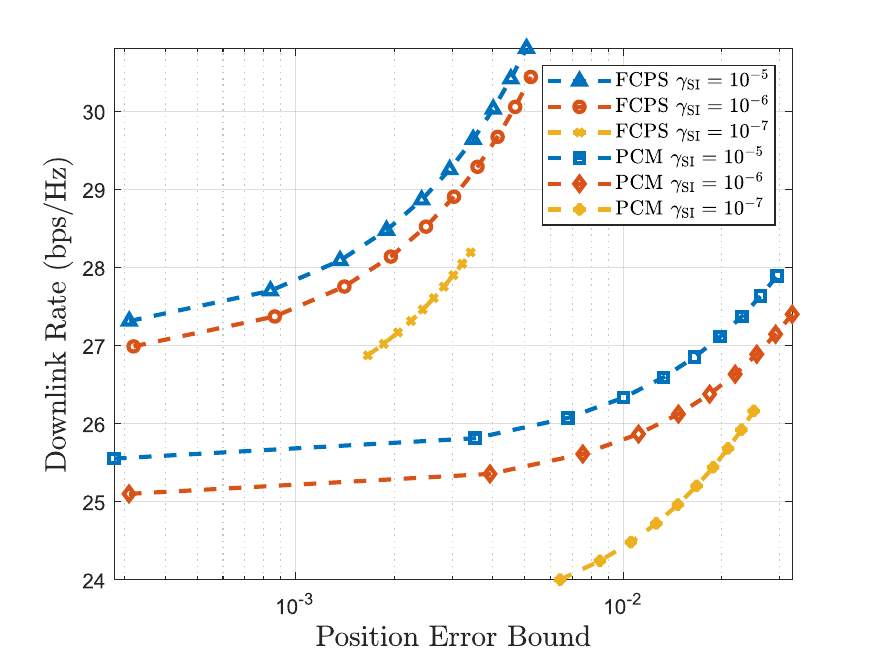}
        \caption{\small{Similar to Fig.~\ref{fig: tradeoff} but for different values for the instantaneous residual SI threshold in the RX of the proposed XL FD node after analog combining, considering both the FCPS and PCM analog BF architectures.}}
        \label{fig: tradeoff_SI}
\end{figure}

\begin{figure*}[t!]
    \centering
    \begin{subfigure}[t]{0.5\textwidth}
        \centering
        \includegraphics[scale=0.55]{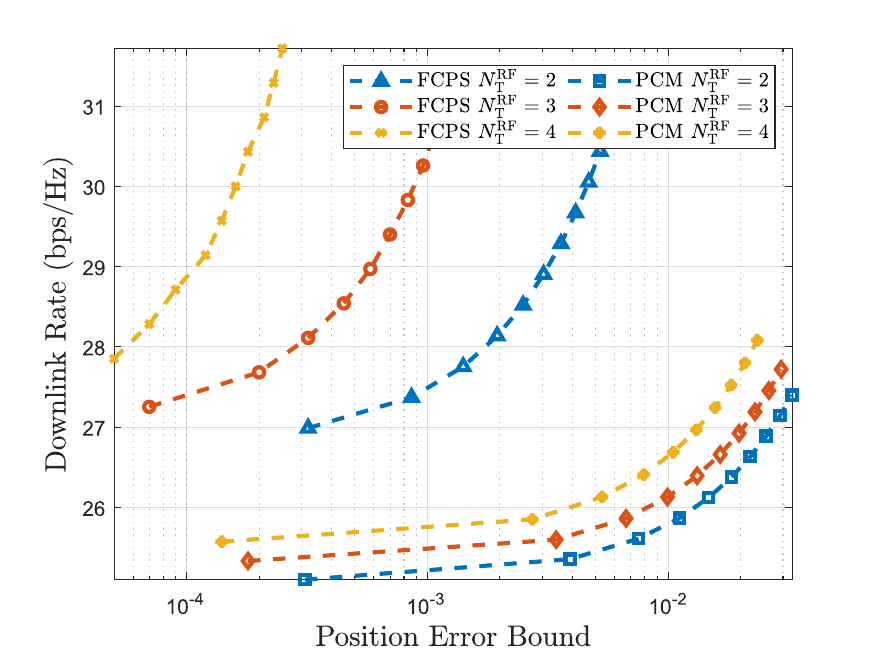}
        \caption{Pareto boundary versus $N_{\rm T}^{\rm RF}$.}
        \label{fig: tradeoff_TX}
    \end{subfigure}%
    \begin{subfigure}[t]{0.5\textwidth}
        \centering
        \includegraphics[scale=0.55]{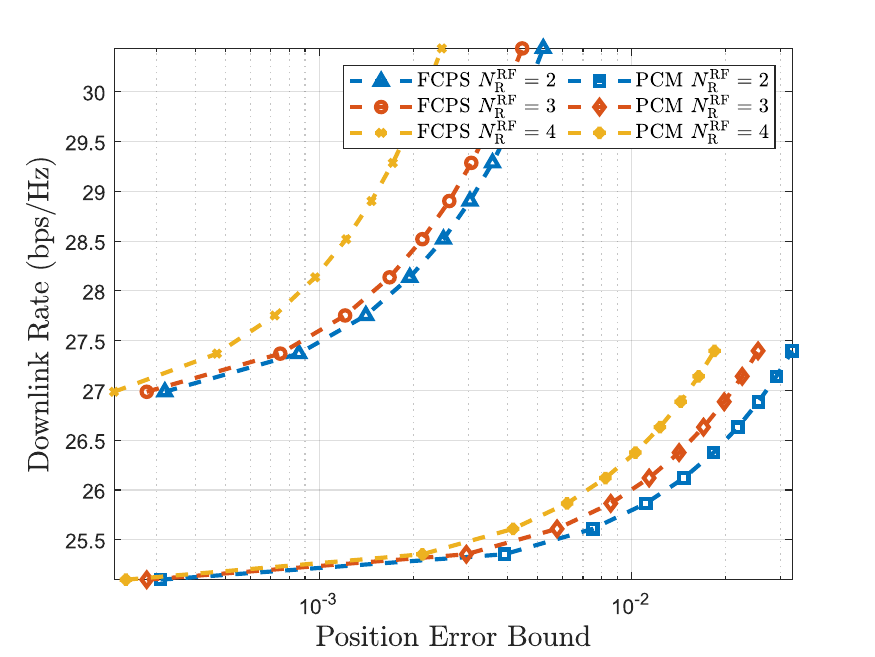}
        \caption{Pareto boundary versus $N_{\rm R}^{\rm RF}$.}
        \label{fig: tradeoff_RX}
    \end{subfigure}
    \caption{\small{Similar to Fig.~\ref{fig: tradeoff} but for different numbers $N_{\rm T}^{\rm RF}$ and $N_{\rm R}^{\rm RF}$ of the TX and RX RF chains, respectively, for the proposed XL FD MIMO system, considering both the FCPS and PCM analog BF architectures.}}
    \label{fig: Tradeoff_TX_RX}
\end{figure*}

The Pareto boundaries resulting from the solution of $\mathcal{OP}_{\rm Pareto}$ for $\nu\in[0,1]$ are illustrated for various configurations of the proposed XL FD MIMO system in Figs.~\ref{fig: tradeoff}--\ref{fig: Tradeoff_TX_RX}. In Figs.~\ref{fig: tradeoff} and~\ref{fig: tradeoff_SI}, we have considered $N_{\rm T}^{\rm RF}=N_{\rm R}^{\rm RF}=2$ TX/RX RF chains, each connected to $N_{\rm E}=128$ antenna elements, both the FCPS and PCM analog BF architectures as well as FDAs, and $P_{\rm{\max}}=36$ dBm. In the former figure, we have set $\gamma_{\rm A}=10^{-6}$, while this threshold has been varied in Fig.~\ref{fig: tradeoff_SI}. In Fig.~\ref{fig: Tradeoff_TX_RX}, we focused on the FCPS and PCM architectures and used a similar setting of the system parameters to Fig.~\ref{fig: tradeoff}, except from varying the numbers $N_{\rm T}^{\rm RF}$ and $N_{\rm R}^{\rm RF}$ of the TX and RX RF chains, respectively. It can be observed from Fig.~\ref{fig: tradeoff} that, for both considered analog BF architectures, the proposed ISAC framework yields improved communications versus sensing trade-off compared to the state-of-the-art design in~\cite{spawc2024}. It is also shown that the performance with the FCPS architecture is almost equivalent with that with FDAs for all $\nu$ values, while, with the PCM architecture, although the achievable DL rate is comparable with the FCPS case, there exists an order of magnitude degradation in the PEB performance. This behavior suggests that, for ISAC applications with less demanding sensing requirements in terms of PEB, the PCM architecture is the most cost- and power-efficient solution.


To investigate the impact of SI cancellation, solely through analog BF, on our XL FD MIMO ISAC designs, we plot in Fig.~\ref{fig: tradeoff_SI} the Pareto boundaries for different $\gamma_{\rm A}$ thresholds. It can be seen that, as $\gamma_{\rm A}$ decreases (i.e., the requirement for SI cancellation becomes less strict), our system's ISAC performance improves for both the FCPS and PCM analog BF architectures. This happens because more system resources can be allocated for ISAC instead for cancelling the SI signals. 
In contrast, when the SI cancellation requirements increase, the performances of both the communication and sensing functionalities degrade. As shown in the figure, the largest ISAC degradation happens for the $\gamma_{\rm A}=10^{-7}$ threshold. Interestingly, for $\gamma_{\rm A}=10^{-6}$, our XL FD MIMO system with both designs continues to operate reasonably well, as the resources required for SI cancellation do not overly constrain its overall efficiency. However, as $\gamma_{\rm A}$ decreases further, a larger portion of the system's effort is allocated to meet the SI cancellation threshold, leaving less capability for ISAC. This results in a significant decline in both communication and sensing performance, with sensing being particularly impacted.


Figures~\ref{fig: tradeoff_TX} and~\ref{fig: tradeoff_RX} demonstrate the Pareto boundaries with for the FCPS and PCM analog BF architectures versus $N_{\rm T}^{\rm RF}$ and $N_{\rm R}^{\rm RF}$, respectively. It can be seen from the former figure that, when $N_{\rm T}^{\rm RF}$ increases, both communication and sensing performances improve, which is attributed to the larger BF gain. In contrast, Fig.~\ref{fig: tradeoff_RX} showcases that, as expected, only the sensing performance improves with increasing $N_{\rm R}^{\rm RF}$, which provides more flexibility in the analog combining design at the RX side of the proposed XL FD MIMO node. Interestingly, it can be observed that similar sensing gains in terms of PEB performance are provided with either larger $N_{\rm T}^{\rm RF}$ or $N_{\rm R}^{\rm RF}$, with the former also contributing in boosting the DL rate performance. It is noted, however, that, when increasing $N_{\rm T}^{\rm RF}$, the SI signal may become stronger, and this needs to be compensated from the proposed ISAC design. 

\section{Conclusions and Future Work}\label{ref:Conclusion}
In this paper, we presented a novel XL FD MIMO framework for simultaneous DL data communications and monostatic-type sensing at sub-THz frequencies, where the latter functionality was based on received reflections of the DL data signals from a passive target lying in the FD node's vicinity either in the near- or far-field regimes. The PEB of the target's 3D spatial parameters considering two analog BF architectures for the TX and RX of the XL FD MIMO node, one based on FCPS and the other on PCM, was analyzed. Two ISAC designs, one per BF architecture, for the TX and RX A/D beamformers as well as the digital SI cancellation matrix at the FD node, with the objective to maximize the achievable DL rate while guaranteeing a maximum PEB threshold, were presented. Our numerical investigations showcased the superiority of both proposed ISAC designs over state-of-the-art schemes, highlighting the potential of the DMA implementation for the proposed XL FD MIMO system which can be easily scaled to larger numbers of metamaterials. In addition, we numerically evaluated the pareto boundary for various system parameters, which unveiled the role of increasing numbers of TX and RX chains and that of the residual SI level in the trade-off between the communication and sensing functionalities for the proposed XL FD MIMO system. For future work, we intend to extend the proposed FD-enabled ISAC framework and estimation error analysis to incorporate multi-UE DL communications, to handle sensing over a candidate geographical area within the FD node's near-/far-field vicinity, rather only around a potential target position, and to more realistic modeling~\cite{williams2022electromagnetic} of the PCM architecture.


\bibliographystyle{IEEEtran}
\bibliography{ms}
\end{document}